\documentclass[12pt,a4paper]{article}
\usepackage{fixltx2e,fix-cm}
\usepackage{amssymb}
\usepackage{amsmath}
\usepackage{graphicx}
\usepackage{makeidx}
\usepackage{multicol}

\makeindex
\usepackage{chicago}
\usepackage{subfig}
\newcommand{\rowgroup}[1]{\hspace{-.5em}#1}

\title{
Inferences on the acquisition of multidrug resistance in \emph{Mycobacterium tuberculosis} using molecular epidemiological data
} 
\author{Guilherme S. Rodrigues\footnote{School of Mathematics and Statistics, University of New South Wales, Australia.}
\hspace{-3mm}
\footnote{CAPES Foundation, Ministry of Education of Brazil, Bras\'ilia - DF 70040-020, Brazil}, Andrew R. Francis\footnote{Centre for Research in Mathematics, Western Sydney University, Australia.},\\
        Scott A. Sisson$^1$ and Mark M. Tanaka\footnote{School of Biotechnology and Biological Sciences, and Evolution \& Ecology Research Centre, University of New South Wales, Australia.}}

\begin{document}
\maketitle

\section{Introduction} 

Tuberculosis (TB) is a lung disease caused by the bacterium {\em Mycobacterium tuberculosis} which kills around 1.5 million people each year and remains a serious challenge for global public health \cite{world2015global}.  Antibiotic drugs for treating TB have been available since the mid 20th century, and currently implemented strategies for TB control rely on the efficacy of these drugs. Treatment of TB involves combination therapy -- in which multiple drugs are administered together in part to improve killing efficacy. The ``first-line'' drugs used in combination to treat tuberculosis are rifampicin, isoniazid, pyrazinamide, ethambutol and streptomycin.

As with most other pathogens, resistance to antibiotic drugs has rapidly evolved in {\em M. tuberculosis}. Streptomycin was the first of the first-line drugs to be developed and deployed in 1943, but resistance was observed before the end of that decade~\cite{Mitchison.51.segregation,Gillespie.02.Evolution}.
Of particular concern is the rise of bacterial strains resistant to multiple drugs, as cases caused by them are difficult to treat successfully. Multidrug resistance (MDR) is defined as resistance to both rifampicin and isoniazid. These are the two most effective drugs against tuberculosis (when the strain is not resistant). Currently, 3.3\% of new TB cases are multi-drug resistant \cite{world2015global}. The occurrence of MDR-TB strains that have additional resistance (called extensively drug resistant, XDR and totally drug resistant, TDR) are particularly problematic and have the potential to cause large outbreaks that are difficult to control \shortcite{Gandhi.06.Extensively}. A better understanding of how multiple drug resistance evolves would aid efforts to contain resistance and control tuberculosis.

Genetic studies have established that many independent mutation events have led to resistance \cite{Ramaswamy.98.Molecular}. Although this suggests that 
mutation of genes is an important source of resistance, model-based analysis of molecular data has revealed that among resistant cases, most are due to the transmission of already resistant bacteria \shortcite{Luciani.09.epidemiological}. 
It is therefore of interest to investigate whether or not this finding also holds for multi-drug resistant tuberculosis.

The rates at which resistance evolves against different drugs vary. 
For instance, isoniazid resistance is known to be acquired faster than rifampicin resistance \shortcite{Ford.13.Mycobacterium,Gillespie.02.Evolution,Nachega.03.Tuberculosis}. 
The rates of mutation to resistance per cell generation are low in absolute value; for example for isoniazid the rate is around $3\times10^{-8}$ and for rifampicin it is  around $2\times10^{-10}$ \cite{David.70.probability,Gillespie.02.Evolution}, although there is a high degree of variation across different lineages of {\em M. tuberculosis} \shortcite{Ford.13.Mycobacterium}. 
One might therefore expect that double resistance of these drugs (MDR) evolves at an exceedingly low rate \cite{Nachega.03.Tuberculosis}. However, MDR strains often occur at appreciable frequencies \shortcite{Zhao.12.National,Anderson.14.Transmission} and a recent study has presented a theoretical model showing how double resistance can evolve rapidly within hosts \shortcite{Colijn.11.Spontaneous}. It would be useful to establish whether such fast direct acquisition of double resistance can be detected in bacterial isolates from epidemiological studies.

\begin{figure}
\begin{center}
\includegraphics[scale=0.8]{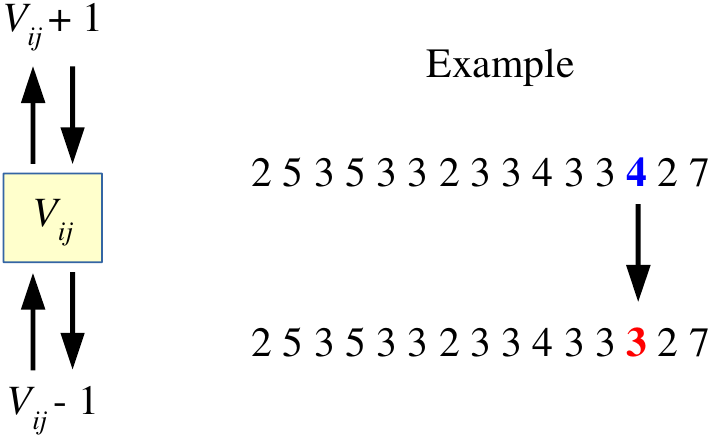}
\caption{\small VNTR loci mutate in a stepwise manner so that the number of repeat units at a locus increases or decreases. In our analysis we assume that when mutation occurs at a locus $j$ in genotype $i$, the repeat number  $V_{ij}$ increases or decreases by a single copy. We further assume that a single unit (repeat number of 1) is an absorbing boundary. The hypothetical example shows how mutation at locus number 13 creates a new VNTR genotype.\label{fig:vntr_mutation}}
\end{center}
\end{figure}

To characterise patterns of TB transmission and drug resistance in a given geographic region, bacterial isolates from TB patients are often genotyped using molecular markers known as variable numbers of tandem repeats (VNTRs) which are repeated genetic sequences that exhibit variation across isolates. The source of this variation is mutation at the VNTR genetic loci which leads to the expansion or contraction of repeat numbers at those loci~(Figure~\ref{fig:vntr_mutation}). A scheme for discriminating effectively among a set of isolates involves considering repeat numbers at multiple VNTR sites. This molecular typing scheme is called multi-locus VNTR analysis (MLVA); in the context of tuberculosis epidemiology it is often known as mycobacterial interspersed repetitive units-VNTR (MIRU-VNTR) \shortcite{Mazars.01.High-resolution,Supply.06.Proposal}.  Typing techniques such as MLVA have been useful for tracking particular strains and understanding how drug resistance evolves and disseminates at the epidemiological level \shortcite{Monteserin.13.genotypes,Anderson.14.Transmission}. 

% ~~~~~~~~~~~~~~~~~~~~~~~~~~~~~~~~~~~~~~~~~~~~~~~~~~~~~~~~~~~~~~~~~~~~
% ~~~~~~~~~~~~~~~~~~~~~~~~~~~~~~~~~~~~~~~~~~~~~~~~~~~~~~~~~~~~~~~~~~~~

Here, we investigate the rates of drug resistance acquisition in a natural population using molecular epidemiological data from Bolivia \shortcite{Monteserin.13.genotypes}. 
First, we study the rate of direct acquisition of double resistance from the double sensitive state within patients and compare it to the rates of evolution to single resistance. 
In particular, we address whether or not double resistance can evolve directly from a double sensitive state within a given host. 
Second, we aim to understand whether the differences in mutation rates to rifampicin and isoniazid resistance translate to the epidemiological scale.  
Third, we estimate the proportion of MDR TB cases that are due to the transmission of MDR strains compared to acquisition of resistance through evolution. 
To address these problems we develop a model of TB transmission in which we track the evolution of resistance to two drugs and the evolution of VNTR loci. 
However, the available data (see Section \ref{sec:data}) is incomplete, in that it is recorded only {for a fraction of the population and} at a single point in time. The likelihood function induced by the proposed model is computationally prohibitive to evaluate and accordingly impractical to work with directly. We therefore approach statistical inference using approximate Bayesian computation techniques.

%%%%%%%%%%%%%%%%%%%%%%%%%%%%
\section{Data}
%%%%%%%%%%%%%%%%%%%%%%%%%%%%
\label{sec:data}

The data set we use is taken from a study of tuberculosis in Bolivia~\shortcite{Monteserin.13.genotypes}. 
Bolivia has a population of 11 million people and a TB incidence of 120 per 100,000 per year. {This rate is comparable to the global incidence of TB (133 per 100,000 per year) and to the rate in Peru, but is 3--6 times the TB incidence in neighbouring countries Brazil, Paraguay, Uruguay, Argentina and Chile~\cite{world2015global}.} 
In the molecular epidemiological study, the investigators genotyped 100 isolates collected in 2010, which represented an estimated 1.1\% of the cases in Bolivia at the time of the study~\shortcite{Monteserin.13.genotypes}. Each isolate was tested for drug sensitivity to five drugs. Here, we focus on resistance against the two drugs isoniazid and rifampicin used to define multidrug resistance. Of the 100 isolates, 14 were found to be MDR, that is, resistant to both of these drugs, 78 were sensitive to both drugs and the remaining 8 were resistant to isoniazid but sensitive to rifampicin. 
No isolates were resistant to rifampicin while being sensitive to isoniazid.

In addition to these drug resistance profiles, each isolate was genotyped using 15 VNTR loci. For example, an isolate in the data set, which was resistant to isoniazid but sensitive to rifampicin, had the following 15 repeat numbers for its 15 VNTR loci: 
$
   143533233433527 
$,
which together constitute its genotype. Variation in these genotypes occurs through a process of mutation in which repeat numbers increase or decrease (see Figure~\ref{fig:vntr_mutation}).

Let $g$ be the number of distinct genotypes present in a sample, 
and label the resistance profiles by ($0$, INH, RIF, MDR), where $0$ denotes sensitivity to both drugs, INH denotes resistance to isoniazid and sensitivity to rifampicin, RIF denotes resistance to rifampicin and sensitivity to isoniazid, and MDR denotes resistance to both drugs. 
The observed data $\mathbf{X}_{obs}$ are then a $g \times 4$ matrix of counts, such that each row gives the distribution of isolates across the four resistance profiles for a given genotype and each column gives the distribution of isolates across genotypes for a given resistance profile.  
The sum of entries in a particular row is the number of isolates with that genotype, while the sum of entries in a particular column is the number of isolates with that resistance profile.
The data set also includes a $g \times 15$ matrix of repeat numbers from the VNTR genotyping.

The Bolivian data set is displayed in full in Table \ref{tab:full_dataste}, which shows all $g=66$ distinct genotypes and classifies all 100 isolates according to genotype and resistance profile. The  $\mathbf{X}_{obs}$ matrix is formed by combining the $0$, INH, RIF and MDR columns.

\begin{table}[ht]
\centering
\scriptsize
\begin{tabular}{lccccclcccc}
\vspace{.1cm}
 {\bf Genotype} & ${\mathbf 0}$ & INH & RIF & MDR & & {\bf Genotype} & ${\mathbf 0}$ & INH & RIF & MDR \\ 
\hline \\[-1mm] 
253533233433427 & 4 & 0 & 0 & 0 & & 243413342212437 & 1 & 0 & 0 & 0 \\ 
253533233433327 & 3 & 0 & 0 & 0 & & 233312442212437 & 1 & 0 & 0 & 0 \\ 
253533233433527 & 11 & 1 & 0 & 0 & & 233313441212437 & 1 & 0 & 0 & 0 \\ 
253533233433525 & 3 & 0 & 0 & 0 & & 233413442212248 & 1 & 0 & 0 & 0 \\ 
143533233433527 & 2 & 1 & 0 & 0 & & 233413442212249 & 1 & 0 & 0 & 0 \\ 
253333244232232 & 1 & 0 & 0 & 0 & & 233213442212349 & 1 & 0 & 0 & 0 \\ 
25333324423-232 & 1 & 0 & 0 & 0 & & 231413542212335 & 1 & 0 & 0 & 0 \\ 
254333243232342 & 0 & 2 & 0 & 2 & & 232433242212436 & 1 & 0 & 0 & 0 \\ 
263532232423139 & 3 & 0 & 0 & 0 & & 234413442212436 & 1 & 0 & 0 & 0 \\ 
223413442212437 & 2 & 0 & 0 & 0 & & 434433452212427 & 1 & 0 & 0 & 0 \\ 
233413542212347 & 0 & 1 & 0 & 2 & & 256432342122237 & 2 & 1 & 0 & 0 \\ 
244333244232332 & 0 & 0 & 0 & 1 & & 256433342123236 & 2 & 0 & 0 & 0 \\ 
244333244232322 & 1 & 0 & 0 & 0 & & 247432342122136 & 1 & 0 & 0 & 0 \\ 
245333244242332 & 1 & 0 & 0 & 0 & & 268432252122227 & 0 & 1 & 0 & 0 \\ 
254333244232232 & 0 & 0 & 0 & 1 & & 268632252122227 & 1 & 0 & 0 & 0 \\ 
254333244232332 & 1 & 0 & 0 & 0 & & 221313352122338 & 0 & 0 & 0 & 1 \\ 
254333244242332 & 1 & 0 & 0 & 0 & & 263532233423148 & 1 & 0 & 0 & 0 \\ 
253333244242232 & 1 & 0 & 0 & 0 & & 360332233423138 & 1 & 0 & 0 & 0 \\ 
252333243232232 & 0 & 0 & 0 & 1 & & 263513233523344 & 1 & 0 & 0 & 0 \\ 
252333243232332 & 0 & 0 & 0 & 1 & & 253523233433527 & 1 & 0 & 0 & 0 \\ 
251333243242332 & 1 & 0 & 0 & 0 & & 253533232433527 & 1 & 0 & 0 & 1 \\ 
252333243262222 & 0 & 0 & 0 & 1 & & 253533232433427 & 1 & 0 & 0 & 0 \\ 
244233234222322 & 1 & 0 & 0 & 0 & & 253523133433527 & 1 & 0 & 0 & 0 \\ 
233373242232325 & 1 & 0 &0 & 0 & & 253533133433527 & 1 & 0 & 0 & 0 \\ 
252343242232524 & 1 & 0 & 0 & 0 & & 353533233433427 & 0 & 0 & 0 & 1 \\ 
25233234423251a & 1 & 0 & 0 & 0 & & 253533233433837 & 1 & 0 & 0 & 0 \\ 
35234234423251a & 1 & 0 & 0 & 0 & & 253533233433237 & 1 & 0 & 0 & 0 \\ 
233413442212338 & 0 & 1 & 0 & 0 & & 254533233433537 & 0 & 0 & 0 & 1 \\ 
233413442212335 & 1 & 0 & 0 & 0 & & 253533233433536 & 1 & 0 & 0 & 0 \\ 
233413442212337 & 1 & 0 & 0 & 0 & & 252533233433428 & 1 & 0 & 0 & 0 \\ 
21341344221233a & 1 & 0 & 0 & 0 & & 253534233433325 & 1 & 0 & 0 & 0 \\ 
213413442212327 & 0 & 0 & 0 & 1 & & 243533232433737 & 1 & 0 & 0 & 0 \\ 
233413442212437 & 3 & 0 & 0 & 0 & & 242433433433436 & 1 & 0 & 0 & 0 \\ 
\end{tabular}
\caption{\small 
Molecular data set compiled from Monteserin et al. (2013). All isolates were classified according to their genotype and resistance profile. The symbol ``a'' represents 10 repeat units and ``-'' represents missing data. 
The entries in the four columns sum to the total number of isolates, 100.
}
\label{tab:full_dataste}
\end{table}

%%%%%%%%%%%%%%%%%%%%%%%%%%%%%%%%  
\section{Model} % (fold) 
%%%%%%%%%%%%%%%%%%%%%%%%%%%%%%%%
\label{sec:description_of_the_model_and_parameters}

In this Section we introduce a model that incorporates both VNTR-based genotyping and drug resistance states. The dynamic variables of the model correspond to numbers of cases of untreated and treated tuberculosis, their resistance states and VNTR genotypes associated with these infections in the population. 
We will now briefly describe processes involved in the model, and provide further details in the following Subsections.

\begin{center}
\begin{figure}[ht]
\includegraphics[width=\textwidth]{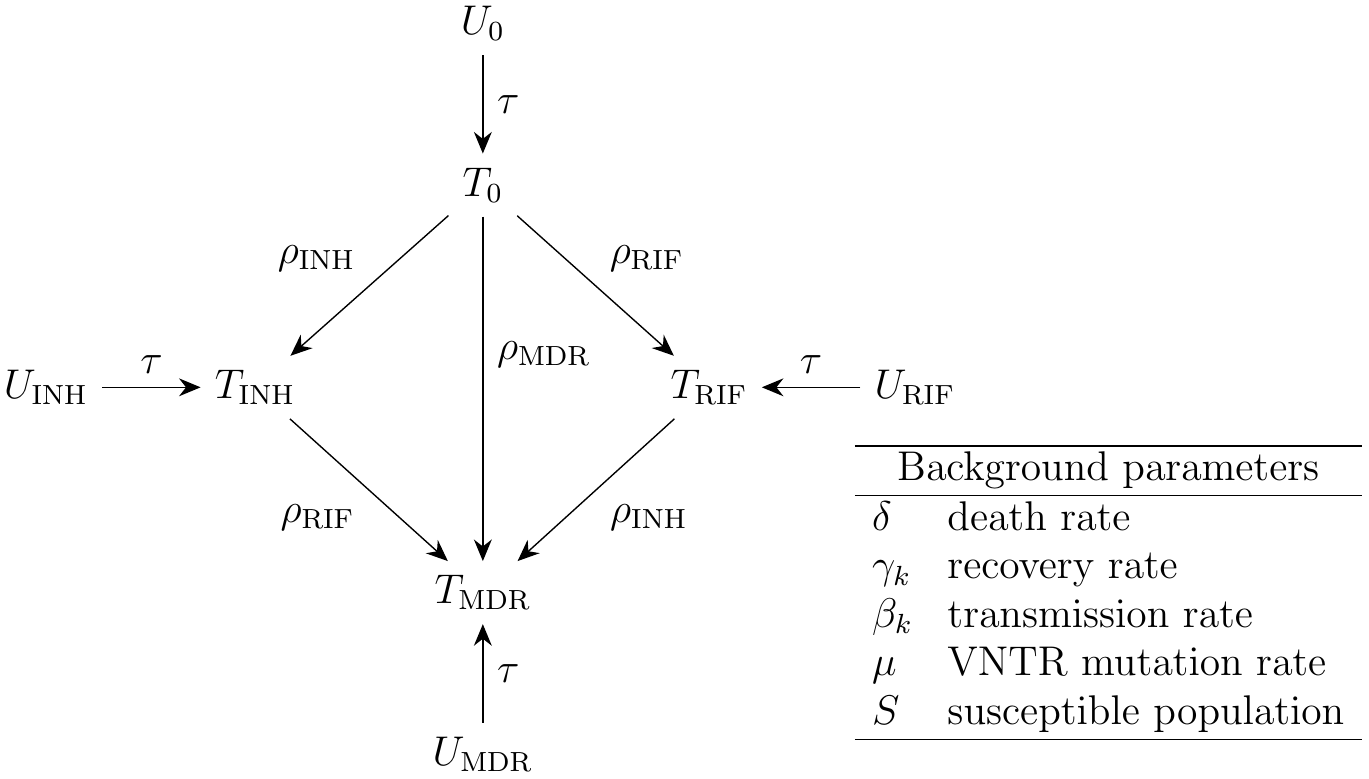} 
\caption{ \small  
Model structure for numbers of untreated ($U_k$) and treated ($T_k$) cases
and per capita rates of conversion (within-host substitution) among resistance classes. Rates are $\rho_{\mathrm{INH}}$ and $\rho_{\mathrm{RIF}}$ for acquisition of resistance to isoniazid and rifampicin respectively, and $\rho_{\mathrm{MDR}}$ for single step acquisition of resistance to both drugs.
Detection (and treatment) of cases is shown with arrows labelled with $\tau$.  
Background parameters are shown in the table to the right, with rates per capita per unit time, and resistance states $k=0$, INH, RIF, MDR.  The mutation process of the VNTR locus is described in 
Section \ref{sec:mutation}. 
\label{fig:modelstructure}}
\end{figure}
\end{center}

An untreated case of TB can become detected and treated, and treatment involves a combination of drugs including the two in question. 
Drug sensitive strains can acquire resistance under treatment with some probability and thereby change their resistance state. 
Treated and untreated cases can infect susceptible individuals and convert them to untreated cases. We disregard latent infections for simplicity (although latency is an important feature of the natural history of tuberculosis), and focus on active infections which are the larger source of new infections. Treated and untreated individuals can also recover or die. Treated individuals enjoy an additional probability of recovery that depends on the efficacy of the drugs, which in turn depends on the sensitivity or resistance of the infecting strain. Treated and untreated cases are also associated with a VNTR genotype, and this genotype evolves over time according to a stepwise mutation process for each locus. Figure~\ref{fig:modelstructure} shows the broad structure of the model with respect to treatment and resistance states, while suppressing details of transmission, recovery, death and mutation of the VNTR loci. 

{At the end of the period of evolution, a simple random sample of 100 isolates is taken without replacement from the population, which matches the sample size of the Bolivian dataset. This provides a full description of the generative process for the observable data.}

Let $G$ be the number of distinct genotypes in the population (the number of distinct genotypes in the \emph{sample} is $g$) and $L$ be the number of VNTR loci used in the genotyping scheme. 
For the Bolivian dataset $L=15$. In the model, the variable $G$ is unknown and varies dynamically. We maintain three matrices which change through time: a $G\times L$ matrix, $\bf V$, which describes the VNTR genotypes; a $G\times 4$ matrix, $\bf U$, which describes the numbers of \emph{untreated} cases of tuberculosis classified according to VNTR genotype and resistance state; and a $G\times 4$ matrix $\bf T$ which describes the numbers of \emph{treated} cases of tuberculosis, again classified according to VNTR genotype and resistance state.
It will be useful to define a $G\times 4$ matrix, $\bf W$, whose entries are the total numbers of both treated and untreated cases: ${\bf W} = {\bf U} + {\bf T}$.

As it will be helpful to be able to pick out columns of these matrices, we adopt notation for the standard basis vectors of $\mathbb R^n$. Let $e_i$ denote the $i$-th basis (column) vector, so that $e_i=(0,\cdots,0,1,0\cdots,0)^{\top}$, with the 1 in the $i$-th position. 
This allows us, for instance, to write the columns of the matrix $\mathbf T$ corresponding to each resistance state 
as $\mathbf T_0=\mathbf T\, e_1$, $\mathbf T_{\mathrm{INH}}=\mathbf T\, e_2$, 
$\mathbf T_{\mathrm{RIF}}=\mathbf T\, e_3$ and $\mathbf T_{\mathrm{MDR}}=\mathbf T\, e_4$, with similar notation for other matrices (note that the dimension of the $e_i$ is left open but inferred from the matrix multiplication; in this case they are in $\mathbb R^4$).

Further, writing $\mathbf 1_i$ for the column vector in $\mathbb R^i$ whose entries are all $1$, then the product $\mathbf T\,\mathbf 1_4$ is a $G\times 1$ column vector whose entries are the numbers of individual cases of each VNTR genotype in the treated population, and the product $\mathbf 1_G^{\top}\,\mathbf T\, \mathbf 1_4$ is the sum of all the entries in $\mathbf T$ (the size of the treated population).  Thus, we can write the size of the susceptible population, $S$, as
\[
   S=N-\mathbf 1_G^{\top}\,\mathbf W\, \mathbf 1_4,
\]
where $\mathbf 1_G^{\top}\,\mathbf W\, \mathbf 1_4$ is the size of the infected population, and where $N$ is the total population size which remains constant. We treat $N$ as modelling the set of all individuals who come in contact with infectious cases and so we exclude individuals who either do not encounter infectious cases or are otherwise protected from infection. This variable therefore may be smaller than the actual population size.

The components of each vector $\mathbf T_k$ for $k=0$, INH, RIF or MDR, are integers, representing the number of individual cases for each genotype.  In the schematic diagram of the model in Figure~\ref{fig:modelstructure}, we use $T_k=\mathbf 1_G^\top\, \mathbf T_k$ to represent the  total population number 
of treated individuals with resistance state $k$, 
with similar notation $U_k=\mathbf 1_G^\top\, \mathbf U_k$ to represent the untreated populations.  
The matrix notation is gathered and shown in Table~\ref{t:matrices}.

The arrows between populations in Figure~\ref{fig:modelstructure} represent the directional rates of detection and treatment $\tau$ and acquisition of resistance to each drug or set of drugs, so that $\rho_{\mathrm{INH}}$ and $\rho_{\mathrm{RIF}}$ represent rates of acquisition of resistance to isoniazid and rifampicin respectively and $\rho_{\mathrm{MDR}}$ the rate of double acquisition.

\begin{table}[ht]
\begin{tabular}{cp{8cm}}
Symbol                                    & Meaning\\
\hline
$\mathbf V$                                 & $G\times L$ matrix describing the VNTR genotypes. \\
$\mathbf U$, $\mathbf T$, $\mathbf W$       & $G\times 4$ matrices of untreated, treated and total cases respectively, with columns corresponding to resistance profiles.\\
$\mathbf U_k$, $\mathbf T_k$, $\mathbf W_k$ & $G\times 1$ column vector for resistance profile $k$ of untreated, treated and total cases respectively.\\
$U_k$, $T_k$, $W_k$                         & Total population sizes of untreated, treated and total with resistance profile $k$.\\ 
$\mathbf U_{i,k}$, $\mathbf T_{i,k}$, $\mathbf W_{i,k}$       & $(i,k)$ entries of the matrices $\mathbf U$, $\mathbf T$, $\mathbf W$: the number of cases in each category with genotype $i$ and resistance profile $k$.\\
$\mathbf 1_i$                               & $i\times 1$ column vector whose entries are all 1.\\
$e_i$                                       & Column vector whose entries are 0 except for 1 in position $i$. Dimension determined by context.\\
\end{tabular}
\caption{\small Summary of linear algebra notation. 
}
\label{t:matrices}
\end{table}

In this model time is discrete, and during each time step the following events takes place in sequence. 
\begin{enumerate}
\item Disease transmission giving rise to new cases; 
\item Natural recovery, cure or death of cases; 
\item Detection of cases which are then treated with drugs;
\item Conversion among resistant profiles in treated cases due to acquisition of resistance; and 
\item Mutation of the genetic marker (multiple VNTR loci).
\end{enumerate}
{The remainder of this Section provides details of how each of these events are modelled. Readers wishing to focus on the statistical aspects of the ABC inference can skip these subsections and go directly to Section~\ref{sec:abc}.}

We regard the above process as a discrete-time stochastic model rather than a discrete-time approximation of a continuous-time stochastic process with rates approximating probabilities, although the latter interpretation becomes more appropriate as the time step length decreases. Here, rates of events will be treated as probabilities, which again is appropriate when time steps are short. The rate parameters are measured in years but we make time steps 1/12 of a year.

A summary of all model parameters, both fixed and to be estimated, and their meanings, are provided in Table~\ref{t:symbols}.

%~~~~~~~~~~~~~~~~~~~~~~~~~~~~~~~~~~~~~~~~~~~~~~~~~~~~~~~~~~~~~~~~~~~~~
%~~~~~~~~~~~~~~~~~~~~~~~~~~~~~~~~~~~~~~~~~~~~~~~~~~~~~~~~~~~~~~~~~~~~~

%~~~~~~~~~~~~~~~~~~~~~~~~~~~~~~~~~~~~~~~~~~~~~~~~~~~~~~~~~~~~~~~~~~~~~
\begin{table}[ht]
\begin{center}
\begin{tabular}{cll}
Symbol & Meaning & Fixed\\
	&	& value\\
\hline
$\delta$ & rate of death and natural recovery & 0.52\\
$\gamma_0$ & cure rate for resistance profile 0, when treated & 0.5 \\ 
$\gamma_{\mathrm{INH}}$ & cure rate for resistance profile INH, when treated & 0.25 \\ 
$\gamma_{\mathrm{RIF}}$ & cure rate for resistance profile RIF, when treated & 0.25 \\ 
$\gamma_{\mathrm{MDR}}$ & cure rate for resistance profile MDR, when treated & 0.05 \\ 
$N$ & total susceptible population size in absence of disease  & $10^4$ \\
$\tau$ & treatment and detection rate &  0.5 \\ 
$c$   & cost of resistance & 0.1 \\
\hline
\vspace{1mm}
Symbol & Meaning & Prior \\
\hline
$\beta_0$ & transmission rate for resistance profile $0$ & Gamma$^*$ \\
$\mu$ & mutation rate of VNTR per locus per unit time & $U(0,1)$ \\
$\rho_\mathrm{INH}$ & rate of acquisition of resistance to INH & $U(0,1)$ \\
$\rho_\mathrm{RIF}$ & rate of acquisition of resistance to RIF  & $U(0,1)$ \\
$\rho_\mathrm{MDR}$ & rate of acquisition of resistance to INH and RIF & $U(0,1)$ 
\end{tabular}
\caption{\small Summary of model parameters.
The top set of parameters are given fixed values, whereas the bottom set of parameters are allocated prior distributions and estimated using ABC. Fixed values and priors are justified in Section~\ref{sec:prior}. Rates are in units of per capita per year, but the time unit is set to 1/12 year in simulations. 
{$^*$Specifically, $\beta_0$ is assumed to follow a (shifted) Gamma prior defined as $\beta_0-0.68 \sim \text{Gamma}(\text{shape}=2, \text{rate}=0.73)$. See Section~\ref{sec:prior} for further details.} 
}\label{t:symbols}
\end{center}
\end{table}
%~~~~~~~~~~~~~~~~~~~~~~~~~~~~~~~~~~~~~~~~~~~~~~~~~~~~~~~~~~~~~~~~~~~~~

%%%%%%%%%%%%%%%%%%%%%%%%%%%%%%%%%%%%%%%%%
\subsection{New infections}
%%%%%%%%%%%%%%%%%%%%%%%%%%%%%%%%%%%%%%%%%
 \label{sub:new_infections}

In our model, new infections occur by mass action. The per capita rate at which a susceptible individual becomes infected by a case with resistance profile $k$ is given by $\beta_k/N$ times the number of infected cases in state $k$. The transmission parameters $\beta_k$ are scaled by $1/N$ for convenience since realistic values of $\beta_k/N$ are typically very small, {and this ensures that the $\beta_k$ are on the natural ``per person per unit time'' scale}.

The acquisition of resistance to antibiotics often comes at a cost to the fitness of the bacterium, and we implement this fitness cost by assuming the transmission rate is lower for cases that carry resistance.  Specifically, we assume a cost of ``$c$ per drug'', so that if $\beta_0$ is the transmission rate of sensitive cases, then cases resistant to one drug transmit at rate $\beta_{\mathrm{INH}}=\beta_{\mathrm{RIF}}=(1-c)\beta_0$ and cases resistant to two drugs at $\beta_{\mathrm{MDR}}=(1-c)^2\beta_0$. Here, $c=0.1$ is considered known and fixed based on previous analyses  of molecular epidemiological data \shortcite{Luciani.09.epidemiological}.

We now construct an expression for the average transmission probability across the infected population.  The matrix $\mathbf W$ records all infected cases with different resistance states in each column, and the individuals corresponding to these columns have different transmission rates $\beta=(\beta_0,\beta_{\mathrm{INH}},\beta_{\mathrm{RIF}},\beta_{\mathrm{MDR}})^\top$.  If we write $D_{\beta}$ for the diagonal matrix whose entries are from $\beta$, then the matrix $\mathbf W\, D_{\beta}$ is the infected population matrix $\mathbf W$ whose columns have been scaled by the entries of $\beta$ (the relevant transmission rates).  The expression
\[
p=\frac{1}{N}\mathbf 1_G^{\mathrm T}\,\mathbf W\, D_{\beta}\,\mathbf 1_4
\]  
then gives the average transmission rate per susceptible individual. 
Since the population size $N$ is usually large and the time steps are short, the 
value for $p$ will nearly always be small. Accordingly, and to ensure that it does not exceed 1, we model the probability of transmission per susceptible individual as $\tilde{p}=\min\{1,p\}$.  

At each time step the number $B$ of new infections is a random variable distributed as 
\[
B\sim \text{Binomial}(S,\tilde{p}).
\]
These $B$ new infections are then allocated across VNTR genotypes and resistance profiles according to the proportions represented by the matrix $\mathbf W\, D_{\beta}$.
That is, a multinomial random sample distributes $B$ according to the existing infected population and their relative transmission rates, so that the resulting allocation is a  $G\times 4$ matrix $\Delta_{\beta}$. 
Finally, as new infections are all assumed to be initially undetected, they are allocated to the untreated subpopulation, so that the matrix $\mathbf U$ is updated to $\mathbf U\rightarrow \mathbf U+\Delta_\beta$.

%%%%%%%%%%%%%%%%%%%%%%%%%%%%%%%%%%%%
\subsection{Cure, recovery and death} 
%%%%%%%%%%%%%%%%%%%%%%%%%%%%%%%%%%%%

Infected individuals who are untreated (the population represented by the counts $\mathbf U$) recover or die at rate $\delta=\delta_r+\delta_d$ per case per time unit, where $\delta_r>0$ is the rate of recovery and $\delta_d>0$ is the rate of death due to any cause. The rate of cure due to successful treatment may vary according to resistance profile, so this rate is given by $\gamma_k$ for $k=0$, INH, RIF, MDR. The number of cures, recoveries and deaths in a time step is given by  
\[
    R \sim \mathrm{Binomial}(U,\delta)
\] 
for the untreated population, where $U=\mathbf{1}_G^\top\,\mathbf{U}\,\mathbf{1}_4$ is the total number of all untreated cases, and
\[
    C_k \sim \mathrm{Binomial}(T_k,\delta+\gamma_k),
\] 
for the treated population, where $T_k$ is the  number of treated individuals with resistance profile $k$ (as defined at the start of this Section). 
The $R$ \emph{untreated} recovered individuals are distributed across both  VNTR genotypes and resistance profiles with a multinomial distribution according to the counts given in $\mathbf U$. These are recorded in the $G\times 4$ update matrix $\Delta_\delta$ (so that the sum of the entries in $\Delta_\delta$ is $R=\mathbf{1}_G^\top\,\Delta_\delta\,\mathbf{1}_4$).
Similarly, the  $C_k$ \emph{treated} recovered individuals of resistance profile $k$ are distributed across the VNTR genotypes according to the distribution observed in $\mathbf{T}_k$. These recovered counts for {\em all} resistance profiles are recorded in the $G\time 4$ update matrix $\Delta_{\delta+\gamma},$ which is constructed from the column vectors of recovered treated counts for profile $k$ in the order $k=0$, INH, RIF and MDR.
The matrices $\mathbf U$ and $\mathbf T$ are then updated to $\mathbf U\rightarrow\mathbf U-\Delta_\delta$ and $\mathbf T\rightarrow \mathbf T-\Delta_{\delta+\gamma}$ respectively.
If the last instance of any genotype is removed by cure, recovery or death, the matrices $\bf U, T, V$ are adjusted by removing the rows corresponding to those genotypes, and the number of genotypes is updated with $G\rightarrow G-1$.

Similarly to the case of new infections (Section \ref{sub:new_infections}),
we assume that the recovery rate due to treatment depends only on the number of drugs the infecting strain is resistant to. Specifically, this implies that $\gamma_{\mathrm{INH}}=\gamma_{\mathrm{RIF}}$.

%%%%%%%%%%%%%%%%%%%%%%%%%%%%%%%%%
\subsection{Detection and treatment} 
%%%%%%%%%%%%%%%%%%%%%%%%%%%%%%%%%

In this model, the detection of cases and the commencement of treatment are combined as a single process. Detected cases are transferred from the untreated class to the treated class. 
We denote this combined detection and treatment rate, per case, per unit time, as $\tau>0$.  With this rate we draw $D$ individuals to transfer between untreated and treated populations, where 
$$D\sim \mathrm{Binomial}(U,\tau).$$
These $D$ individuals are then allocated across VNTR genotypes 
and resistance profiles according to the observed distribution of untreated cases, $\mathbf U$.  As before, this results in a $G\times4$  update matrix $\Delta_\tau$, which we use to update
$\mathbf U\rightarrow\mathbf U-\Delta_\tau$ and $\mathbf T\rightarrow\mathbf T+\Delta_\tau$.

%%%%%%%%%%%%%%%%%%%%%%%%%%%%%%%%%%%%
\subsection{Acquisition of drug resistance} 
%%%%%%%%%%%%%%%%%%%%%%%%%%%%%%%%%%%%

Individual treated cases are able to convert from one resistance profile to another through adaptive evolution. 
That is, under drug treatment, natural selection acts to favour increasing levels of resistance.  As a result of this process, individuals may move from the $k=0$ resistance profile 
(sensitive to both drugs) to one of the other three resistance profiles: 
${\mathrm{INH}}$, ${\mathrm{RIF}}$, or ${\mathrm{MDR}}$ 
(resistance to one or both drugs).  Individuals may also move from resistance to exactly one of the drugs (${\mathrm{INH}}$ or ${\mathrm{RIF}}$) to the multiple drug resistance profile ${\mathrm{MDR}}$.  We respectively denote the rate of acquisition of resistance to INH or RIF  by $\rho_{\mathrm{INH}}$ and $\rho_{\mathrm{RIF}}$, and denote the rate of acquisition of resistance from individuals in the sensitive population to both drugs simultaneously by $\rho_{\mathrm{MDR}}$.  These conversions and rates are illustrated schematically in Figure~\ref{fig:modelstructure}.

To model resistance acquisition, we select individuals to move between resistance profiles in  the treated population i.e. between columns in the matrix $\mathbf T$. Acquiring resistance to the drug rifampicin will result in individuals moving from the column 
$\mathbf T_0$ to $\mathbf T_{\mathrm{RIF}}$, and from $\mathbf T_{\mathrm{INH}}$ to $\mathbf T_{\mathrm{MDR}}$ 
at a rate $\rho_{\mathrm{RIF}}$.
Similarly, acquiring resistance to the drug isoniazid results in individuals moving from the column 
$\mathbf T_0$ to $\mathbf T_{\mathrm{INH}}$, and from $\mathbf T_{\mathrm{RIF}}$ to $\mathbf T_{\mathrm{MDR}}$
at a rate $\rho_{\mathrm{RIF}}$. 
Simultaneous acquisition of resistance to both drugs moves individuals from the column $\mathbf T_{0}$ to $\mathbf T_{\mathrm{MDR}}$ at the rate $\rho_{\mathrm{MDR}}$.
These movements occur between columns but not across rows (infections do not change VNTR genotypes through this process).

Mechanistically, we can obtain the number of cases of genotype $i$ transitioning from resistance profile $k$ to resistance profile $k'$, denoted $A_{i,k\rightarrow k'}$, as
\begin{align*}
A_{i,0\rightarrow *}& \sim \mathrm{Multinomial}(\mathbf T_{i,0},\rho_{0\rightarrow *})\\
A_{i,\mathrm{INH}\rightarrow \mathrm{MDR}} & \sim\mathrm{Binomial}(\mathbf T_{i,\mathrm{INH}},\rho_{\mathrm{RIF}})\\
A_{i,\mathrm{RIF}\rightarrow\mathrm{MDR}} & \sim\mathrm{Binomial}(\mathbf T_{i,\mathrm{RIF}},\rho_{\mathrm{INH}})
\end{align*}
where $A_{i,0\rightarrow *}=(A_{i,0\rightarrow\mathrm{INH}}, A_{i,0\rightarrow\mathrm{RIF}}, A_{i,0\rightarrow\mathrm{MDR}}, A_{i,0\rightarrow 0})^\top$ is the vector of cases transitioning from sensitivity, $\mathbf T_{i,k}$ is the entry of the matrix $\mathbf T$ corresponding to the genotype $i$ and resistance profile $k$ (see Table~\ref{t:matrices}), 
 and
$\rho_{0\rightarrow *}=(\rho_{\mathrm{INH}}, \rho_{\mathrm{RIF}},\rho_{\mathrm{MDR}}, 1-\sum_k \rho_k)^\top$
is the vector of probabilities of these events.

If we denote $\Delta_{k\rightarrow k'}$ as column vectors of counts of movements from resistance profile $k$ to $k'$ across all $G$ genotypes, we can then construct the overall $G\times 4$ update matrix
\[
\Delta_\rho = \left(
	\Delta_{0} \mid
	\Delta_{\mathrm{INH}} \mid
	\Delta_{\mathrm{RIF}} \mid
	\Delta_{\mathrm{MDR}}
	\right).
\]
from the column vectors $\Delta_k$, which denote the total population change for resistance profile $k$, where
\begin{align*}
\Delta_{0}  &= -(\Delta_{0\rightarrow\mathrm{INH}}+\Delta_{0\rightarrow\mathrm{RIF}}+\Delta_{0\rightarrow\mathrm{MDR}}) \\
\Delta_{\mathrm{INH}}&= \Delta_{0\rightarrow\mathrm{INH}}- 
\Delta_{\mathrm{RIF}\rightarrow\mathrm{MDR}}
\\
\Delta_{\mathrm{RIF}}&=	\Delta_{0\rightarrow\mathrm{RIF}}
\Delta_{\mathrm{INH}\rightarrow\mathrm{MDR}}
 \\
\Delta_{\mathrm{MDR}}&=	\Delta_{0\rightarrow\mathrm{MDR}}+\Delta_{\mathrm{RIF}\rightarrow\mathrm{MDR}}+\Delta_{\mathrm{INH}\rightarrow\mathrm{MDR}}.
\end{align*}
The population of treated cases is then  updated to $\mathbf T\rightarrow\mathbf T+\Delta_\rho$.

%%%%%%%%%%%%%%%%%%%%%%%%%%%%%%%%%%%%
\subsection{Mutation of the marker} 
%%%%%%%%%%%%%%%%%%%%%%%%%%%%%%%%%%%%
\label{sec:mutation}

The set of $L=15$ VNTR loci constitute the genetic marker used to genotype bacterial isolates (see Section~\ref{sec:data}). Each genotype is a list of numbers of tandem repeat units at the $L$ loci.  
The states of all VNTRs in the infected population are given by the $G\times L$ matrix $\bf V$ with elements $V_{ij}$ describing the repeat number of locus $j$ in genotype $i$.  
Each locus mutates through a stepwise mutation process at rate $\mu$ per locus per case per unit time. When mutation occurs, the repeat number $V_{ij}$ at a locus $j$ of genotype $i$ 
changes by $+1$ or $-1$, each with probability 0.5. A repeat number of 1 is treated as an absorbing boundary (i.e. there is zero probability of the repeat number increasing from 1 to 2) because at state 1 there is no longer a genetic sequence that is tandemly repeated and no mechanism such as replication slippage acts to expand it from 1 to 2. 

Mutation of the marker has the effect of moving cases between the rows of the matrix $\mathbf W$.   
We first identify the number of mutation events in the population, $M$, where $M\sim\mathrm{Binomial}(S,\mu)$ and $S=N-\mathbf 1_G^{\mathrm T}\,\mathbf W\, \mathbf 1_4$ is the size of the susceptible population (see Section~\ref{sec:description_of_the_model_and_parameters}).  The $M$ cases are then distributed across the population of VNTR genotypes and resistance profiles, according to the entries of the matrices $\mathbf T$ and $\mathbf U$.  
Each individual case undergoing mutation corresponds to a specific entry in either $\mathbf T$ or $\mathbf U$.  This entry is described by its VNTR genotype $\mathbf V_i=(V_{i,1},\ldots,V_{i,L})$ where $L=15$ for the Bolivian data, and its resistance profile, $k=0$, INH, RIF, MDR.  The result of the mutation is a change to the VNTR genotype, which is represented by a change in the repeat number at a single locus, $V_{ij}$, by $\pm 1$.  This may or may not result in a VNTR genotype that is already present in the population.

If the new VNTR genotype already appears as a row in the matrix $\mathbf V$ as an existing type in the data, then there is no change to $\mathbf V$.  The matrix $\mathbf T$ or $\mathbf U$ on the other hand is changed by subtracting 1 from one entry and adding one to another entry in the same column (the resistance profile, $k$,  does not change).  In matrix terms, supposing the change is to a treated case, this can be described by updating $\mathbf T\rightarrow\mathbf T-e_{i,j}+e_{i,k}$, where $e_{i,j}$ is the matrix whose entries are zero except for a 1 in the $(i,j)$-th position, and where the VNTR genotype changes from row $j$ to row $k$.

If the new VNTR genotype does not already appear in the population, then the matrix $\mathbf V$ is expanded to include a new row describing the new genotype, so that $\mathbf V$ becomes a $(G+1)\times 15$ matrix.  The update for $\mathbf T$ or $\mathbf U$ is the same as described above, except that now both matrices are $(G+1)\times 4$ dimensional. Subsequent to this update we increment $G\rightarrow G+1$.
If mutation of a VNTR genotype removes the last instance of the original genotype from $\bf U$ and $\bf T$ the corresponding rows of matrices $\bf V$, $\bf U$ and $\bf T$ are deleted, requiring the update $G\rightarrow G-1$.

%%%%%%%%%%%%%%%%%%%%%%%%%%%%%%%%%%%%%
\subsection{Initial conditions of the model} 
\label{sec:initcond}
%%%%%%%%%%%%%%%%%%%%%%%%%%%%%%%%%%%%%

The model covers the period from when drugs are introduced at time $t=0$ to when sampling occurs. 
Since the main first-line anti-tuberculosis drugs were discovered/developed in the 1940s to early 1960s, we assumed treatment commenced around 1960 and ran the simulation for a period of 50 years.  We assumed that both drugs, isoniazid and rifampicin, were introduced at the same time and are administered together in combination therapy.  The standard course of treatment includes both drugs along with other first-line drugs \cite{world2015global}. 

We assume that at the start of the process all cases are sensitive to both drugs and that the number of cases is at equilibrium in the absence of treatment and resistance. 
To compute this equilibrium state, we consider the differential equation describing the deterministic version of the model ignoring VNTR genotypes. Namely, 
\begin{align*}
   \frac{dU}{dt} &= (\beta/N) S U - \delta U 
\end{align*}
where $S=N-U$ and $t$ indicates time. 
Setting $dU/dt$ to zero and solving for the dynamic variables we obtain equilibrium values of 
\[
  \hat{U} = N 
      \left( 
      1 - \frac{\delta}{\beta_0}
     \right) 
     \qquad\mbox{ and }\qquad
     \hat{S} = \frac{\delta N}{\beta}
\]
for $U>0$.

The basic reproduction number of a pathogen $R_0$ is defined to be the average number of new infectious cases caused by a single infection in a completely susceptible population. In our model, before there is any treatment, assuming all cases are doubly susceptible, a single case on average persists for $1/\delta$ years and generates $S\beta_0/N$ new cases per unit time but since $S=N$ in a wholly susceptible population then  $R_0=\beta_0/\delta$.

All cases are initially untreated and sensitive. From time $t=0$ treatment in the population commences. 
To reintroduce into the model genetic variation at the marker loci, the initial distribution of genotype clusters is a random sample drawn from the infinite alleles model from population genetic theory
\shortcite{Ewens.72.Sampling,Hubbell.01.unified,Luciani.08.Interpreting}. The infinite alleles model depends on a single parameter, the diversity parameter, which we set to $2\hat{U} \mu L$ where $\hat{U}$ is the number of cases, taken from the equilibrium value described above, $\mu$ is the mutation rate per VNTR locus and $L$ is the number of VNTR loci used in genotyping isolates. 
To initialise the multi-locus VNTR genotypes, each genotype  is a sequence of random integers, of length $L$, with each VNTR number $V_{ij}$ drawn from a discrete uniform distribution over $\{1,\ldots ,10\}$. 
{
Although the initial distribution of genotype clusters is set under the infinite alleles model, the mutation process for VNTRs brings the distribution in line with the stepwise model over time. 
}

The initial conditions are a function of the parameters which are set according to the priors specified in Section \ref{sec:prior}.

%%%%%%%%%%%%%%%%%%%%%%%%%%%%%%%%%%%%%
\section{Inference with approximate Bayesian computation} % (fold)
%%%%%%%%%%%%%%%%%%%%%%%%%%%%%%%%%%%%%
\label{sec:abc}

For the model in Section \ref{sec:description_of_the_model_and_parameters}, when the data are only observed at a single point in time, the cost of evaluating the likelihood function is computationally prohibitive. This results from the ``incomplete'' nature of the observed data (see Section \ref{sec:data}) in the sense that we only have access to a snapshot of the population, via the observed sample,  at the time the study was conducted, with no direct measurements of the system as it progressed. Computing the likelihood then requires integrating over all potential trajectories the population could have gone through before reaching its final, observed state. 

As such we adopt approximate Bayesian computation (ABC) methods as a means of performing Bayesian statistical inference for the unknown model parameters $\theta=(\beta_0,\mu,\rho_\mathrm{ INH},\rho_\mathrm{RIF},\rho_\mathrm{MDR})^\top$. 
As observed in other chapters in this Handbook, the ABC approximation to the true posterior distribution is given by
\[
	\pi_{ABC}(\theta|s_{obs}) \propto \pi(\theta)\int K_h(\|s-s_{obs}\|)p(s|\theta)ds,
\]
where $\pi(\theta)$ is the prior distribution, $s=S(\mathbf X)$ is a vector of summary statistics with $s_{obs}=S(\mathbf X_{obs})$, $p(s|\theta)$ is the computationally intractable likelihood function for the summary statistics $s$, and $K_h(u)=K(u/h)/h$ is a standard smoothing kernel with scale parameter $h>0$. In the following analyses we used the uniform kernel on $[-h,h]$ for $K_h(u)$. The quality of the ABC approximation depends on the information loss in the summary statistics $s$ over the full dataset $\mathbf  X$, and the size of the kernel scale parameter $h$ with smaller $h$ producing greater accuracy and increased computational cost. Choice of both $s$ and $h$ are typically driven by the amount of expert knowledge and computation available for the analysis. 

For the present analysis we implement a version of a simple ABC importance sampling algorithm, as outlined in the box. Given a suitable importance sampling distribution $q(\theta)$, the algorithm produces a set of weighted samples from the ABC approximation to the true posterior $(\theta^{(1)},w^{(1)}),\ldots,(\theta^{(\tilde{N})},w^{(\tilde{N})})\sim\pi_{ABC}(\theta|s_{obs})$.
As with standard importance sampling, suitable choice of $q(\theta)$ is important to avoid high variance in the importance weights, and also to avoid needlessly generating datasets $s=S(\mathbf X^{(i)})$, $\mathbf X^{(i)}\sim p(\mathbf  X|\theta)$ for which $s^{(i)}$ and $s_{obs}$ will never be close.

\medskip
\noindent
\fbox{
\begin{minipage}{\textwidth}
\begin{center}
\textbf{ABC Importance Sampling Algorithm}
\end{center}
\vspace{-5mm}

\noindent {\it Inputs:}
\begin{itemize}
\item A target posterior density $\pi(\theta|\mathbf X_{obs})\propto p(\mathbf  X_{obs}|\theta)\pi(\theta)${, consisting of a prior distribution $\pi(\theta)$ and a procedure for generating data under the model  $p(\mathbf X_{obs}|\theta)$.}
\item A proposal density $q(\theta)$, with $q(\theta)>0$ if $\pi(\theta|\mathbf X_{obs})>0$.
\item An integer $\tilde{N}>0$.
\item An observed vector of summary statistics $s_{obs}=S(\mathbf  X_{obs})$.
\item A kernel function $K_h(u)$ and scale parameter $h>0$.
\end{itemize}

\noindent {\it Sampling:}\\
\noindent For $i=1, \ldots, \tilde{N}$:
\begin{enumerate}
\item \label{rejection:step1} Generate $\theta^{(i)}\sim q(\theta)$ from sampling density $q$.
\item Generate $\mathbf  X^{(i)}\sim p(\mathbf X|\theta^{(i)})$ from the likelihood.
\item Compute the summary statistics $s^{(i)}=S(\mathbf  X^{(i)})$.
\item Assign $\theta^{(i)}$ the weight $w^{(i)}\propto K_h(\|s^{(i)}-s_{obs}\|)\pi(\theta^{(i)})/q(\theta^{(i)})$.
\end{enumerate}

\noindent {\it Output:}\\
A set of weighted parameter vectors $\{(\theta^{(i)},w^{(i)})\}_{i=1}^{\tilde{N}}$ $\sim$ $\pi_{ABC}(\theta|s_{obs})$.
\end{minipage}
}
\medskip

To determine a suitable importance sampling distribution $q(\theta)$ we adopt a two stage procedure,  following the approach of \citeN{fearnhead+p12}. In the first stage we  perform a pilot ABC analysis using a sampling distribution that is diffuse enough to easily encompass the ABC posterior approximation obtained for a moderate value of the kernel scale parameter $h$. 
We specified $q(\theta)\propto\pi(\theta)I(\theta\in A)$ which is proportional to the prior, but restricted to the hyper-rectangle $A$. Here, $A$ is constructed as the 
smallest credible hyper-rectangle that we believe contains the ABC posterior approximation.
As such, this $q(\theta)$ will identify the general region in which $\pi_{ABC}(\theta|s_{obs})$ is located.
Specifically, for $\theta=(\beta_0,\mu,\rho_{\mathrm{INH}},\rho_{\mathrm{RIF}},\rho_{\mathrm{MDR}})^\top$ we adopt 
$q(\theta) =   %(delta'=.68, 15)
\tilde{\pi}_{15}(\beta_0)\times U(0, .005)\times U(0, .01)\times U(0, .005)\times U(0, .001)$, where $\tilde{\pi}_{15}(\beta_0)$ is the prior $\pi(\beta_0)$ for $\beta_0$ specified in Section \ref{sec:prior}, but truncated to exclude density above the point $\beta_0=15$.

For posterior distributions with strong dependence between parameters, defining $q(\theta)$ over such a hyper-rectangle may be inefficient as it will cover many regions of effectively zero posterior density. 
Accordingly we construct the sampling distribution for the second stage, with the lowest value of $h$, as a kernel density estimate of the previous ABC estimate of the posterior distribution: $q(\theta)=\sum_i w^{(i)}L(\theta|\theta^{(i)})$, where $L$ is a suitable kernel density (not to be confused with the kernel $K_h$). This approach follows the ideas behind the sequential Monte Carlo-based ABC samplers of \shortciteN{Scott2007} and others. 
At each stage the kernel scale parameter $h$ is decreased, and determined as the value which results in $\sim$ 2,000 posterior samples with non-zero weights, for the given computational budget.

To  ensure greater efficiency at each stage
we also performed a non-linear regression adjustment using a neural network with a single hidden layer (see \shortciteNP{blum+f10,abc.package,Beaumont2002}), as implemented in the  {\em R} package {\tt abc}. The adjustment used logistic transformations for the response.

For samples drawn from the final importance sampling distribution $q(\theta)$, the data generation procedure took on average $\sim40$ seconds in {\em R}. {This is computationally expensive from an ABC context, and could be reduced by recoding the simulator in a compiled language such as {\em C}, or by adapting the ``lazy ABC'' ideas of \shortciteN{prangle16lazy} to terminate early those simulations that are likely to be rejected. In this implementation we performed importance sampling from each distribution $q(\theta)$ in parallel on multiple nodes of a computational cluster.}

%%%%%%%%%%%%%%%%%%%%%%%%%%%%%%%%%%% 
\subsection{Summary statistics} 
%%%%%%%%%%%%%%%%%%%%%%%%%%%%%%%%%%%
\label{subsec:summary_statistics}

Considering the matrix structure of the observed  data $\mathbf X_{obs}$ (see Section \ref{sec:data}), we determine the information content in $\mathbf{X}$ as if it was the design matrix of a regression model and summarise it accordingly. Specifically, we define the summary statistics  $s=S(\mathbf X)$ to be the upper-triangular elements of the matrix 
\[
 (\mathbf 1_g\vert  \mathbf{X})^\top (\mathbf 1_g \vert \mathbf{X}),
\]
where the vertical lines denote the addition of an extra column.
The added columns of ones enriches the set of summary statistics by including the row and column totals of $\mathbf X$.
Alternatively, these summary statistics can be described as:
\begin{enumerate}
\item[i)] $g$: the number of distinct genotypes in the sample. 
\item[ii)] $n_k$: the number of isolates with resistance profile $k=$ $0$, INH, RIF and MDR. % 
\item[iii)] $c_{k, k'}=(\mathbf{X}_k)^\top \, \mathbf{X}_{k'}$: the dot product between the resistance profiles of $k$ and $k'$ within  $\mathbf{X}$. 
\end{enumerate}

Note that these summary statistics are over specified in that $n_0+n_\mathrm{INH}+n_\mathrm{RIF} + n_\mathrm{MDR}$ equals the total number of isolates sampled from the population, which is known and equal to the number of isolates in the observed data sample (100 for the Bolivian data). Accordingly, and without loss of generality, we remove $n_\mathrm{MDR}$ as a summary statistic to avoid collinearity.
In combination, this set of 14 summary statistics efficiently encapsulates the available information about the covariance structure of the original dataset $\mathbf{X}$, the distribution of the isolates among the different resistance profiles and the degree of diversity of isolates within the sample.

For the Bolivian dataset, there are $g=68$ distinct genotypes,  $n_{0}=78$ sensitive isolates, $n_{\text{INH}}=8$ isolates resistant to isoniazid only, $n_{\text{RIF}}=0$ isolates resistant to rifampicin only and $n_{\text{MDR}}=16$ doubly resistant isolates (see Table~\ref{tab:full_dataste}). The remaining statistics, $c_{k, k'}$, are computed as:
\begin{center}
\begin{tabular}{lcccc}
      & 0 & INH & RIF & MDR \\
  0   & 232 & 15 & 0 & 1 \\
  INH & -- & 10 & 0 & 6 \\
  RIF & -- & -- & 0 & 0 \\
  MDR & -- & -- & -- & 18 \\
\end{tabular}
\end{center}

Finally, in order to reduce the impact of summary statistics operating on different scales, we compare simulated and observed summary statistics within the kernel $K_h(\|s-s_{obs}\|)$ via the $L_{\frac{1}{2}}$ norm
\[
	\|s-s_{obs}\|=\|S(\mathbf X)-S(\mathbf X_{obs})\| = 
	\left(\sum_{j=1}^{\dim(s)}\left[ S(\mathbf X)_j - S(\mathbf X_{obs})_j\right]^{\frac{1}{2}}\right)^2,
\]
where $\dim(s)=14$ is the number of summary statistics. Alternative approaches could rescale the statistics via an appropriate covariance matrix (e.g. \shortciteNP{Luciani.09.epidemiological,erhardt+s16}) {or use other norms},  however the results in the following Section proved to be robust to more structured comparisons, so we did not pursue this further.
{In particular the following results were robust to these choices because of the use of a good (non-linear) regression adjustment, which greatly improves the ABC posterior approximation, and which has a larger impact on this approximation than the choice of metric $\|\cdot\|$.}

%%%%%%%%%%%%%%%%%%%%%%%%%%%%%%%%%%%%%%%%%%
\subsection{Parameter specifications and prior distributions } 
%%%%%%%%%%%%%%%%%%%%%%%%%%%%%%%%%%%%%%%%%%
\label{sec:prior} 

Of the 13 model parameters (see Table \ref{t:symbols}), eight of these are known well enough for the purposes of our analysis to fix their values. Namely, the parameters $(\delta,\gamma_0,\gamma_\mathrm{ INH},\gamma_\mathrm{RIF},\gamma_\mathrm{MDR}, N, \tau, c)^\top$
are set to these fixed values.  We justify our choices for these values below. The remaining five parameters 
$\theta=(\beta_0,\mu,\rho_\mathrm{ INH},\rho_\mathrm{RIF},\rho_\mathrm{MDR})^\top$ 
are to be estimated, and require a prior distribution specification.

The rate of death or recovery, $\delta$, is fixed and set to be $\delta=0.52$ per case per year following \citeN{Dye.01.Will} and \citeN{Cohen.04.Modeling}. 
Similarly, following \citeN{Dye.01.Will}, untreated individuals are detected and treated at rate $\tau=0.5$ per case per year. 
The rates of recovery due to treatment, $\gamma_k$, for resistance profiles $k=0$, INH, RIF and  MDR, can be written in terms of the probability of treatment success 
\[
    p_k = \frac{\delta_r+\gamma_k}{\delta_d+\delta_r+\gamma_k}.
\]
We set the cure rates to be  $\gamma_0=0.5, \gamma_{\rm INH}=\gamma_{\rm RIF}=0.25,  \gamma_{\rm MDR}=0.05$, which, by using $\delta_r=0.2$ \shortcite{Dye.01.Will,Cohen.04.Modeling}, corresponds to treatment success probabilities of approximately $p_{\rm 0}=0.69$, $p_{\rm INH}=p_{\rm RIF}=0.58$ and  $p_{\rm MDR}=0.44$. These values are within the supported ranges in the literature, namely,  $p_{\rm 0}=0.45-0.75$, $p_{\rm INH}=p_{\rm RIF}=0.3-0.6$ and  $p_{\rm MDR}=0.05-45$ \cite{Blower.04.Modeling}. We chose higher values within these ranges since  \citeN{Blower.04.Modeling} explored a wide range of possibilities in models  including epidemiologically pessimistic scenarios.

The fitness cost of drug resistance, $c$, was fixed and set to be $c=0.1$ based on estimates by \shortciteN{Luciani.09.epidemiological}. 
To set the total population size $N$ we first observe that because the sample of 100 isolates represents $\sim1.1\%$ of the population, this implies that the infected population is 9091. We expect that the number of susceptible individuals who are exposed to disease is somewhat higher than this. Accordingly, we assumed that the total size of the population susceptible to tuberculosis is $N=10,000$. 
Larger total population sizes can be used, at the price of greater computational overheads for generating data under the model.

Previous work estimated rates of resistance acquisition by mutation to be around $0.0025-0.02$ per case per year \shortcite{Luciani.09.epidemiological}. The rate of mutation of the VNTR loci in {\em M. tuberculosis} was estimated to be around  $10^{-3}$ per locus per case per year \shortcite{Reyes.10.Mutation,Aandahl.12.model,Ragheb.13.mutation} but lower estimates have also been found \shortcite{Wirth.08.Origin,Supply.11.mutation}. All of these mutation rates are much lower than 1. We treat these mutation rate parameters as probabilities and conservatively set the standard uniform distribution as a wide prior on each parameter. 
That is, for the acquisition of resistance to isoniazid or rifampicin (or both), we specify priors for the rates of resistance acquisition as $\rho_{\mathrm{INH}},\rho_{\mathrm{RIF}},\rho_{\mathrm{MDR}}\sim U(0,1)$.
Similarly, for the mutation rate of the VNTR molecular marker, $\mu$,  we use the prior $\mu\sim U(0,1)$.

The transmission parameter for doubly sensitive strains $\beta_0$ is given the shifted gamma prior
\[
  \beta_0-0.68 \sim \text{Gamma}(\text{shape}=2, \text{rate}=0.73)
\] 
where the parameters are chosen such that the resulting prior distribution of the basic reproduction number $R_0$ closely resembles the distribution obtained in a numerical analysis of tuberculosis dynamics by \shortciteN{Blower.95.intrinsic}. Note that the prior on $\beta_0$ is shifted in order ensure the realistic condition that $R_0>1$. A value of $R_0$ lower than unity would lead to extinction of {\em M. tuberculosis}.

We reiterate that we interpret the rate parameters as probabilities per time step and handle the parameters so that their values remain in (0,1). This approximation increases in accuracy as the time unit decreases. Here we divide the natural time unit of one year into new units of 1/12 year per time step.

%%%%%%%%%%%%%%%%%%%%%%%%%%%%%%%%%%%%%%
\section{Competing models of resistance acquisition} 
%%%%%%%%%%%%%%%%%%%%%%%%%%%%%%%%%%%%%%

\begin{figure}[ht]
\begin{center}
\includegraphics[scale=0.65]{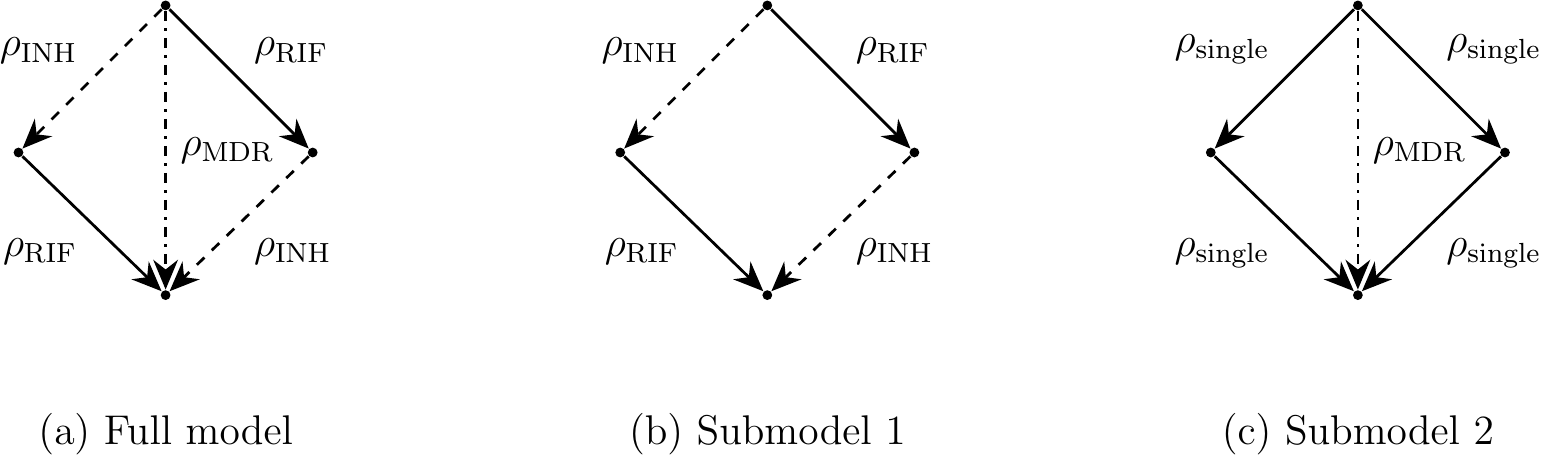}
\end{center}
\caption{\small Three candidate models of acquisition of multiple drug resistance.
(a) The full model: two different rates of conversion leading to acquisition of resistance and a rate of conversion from resistance profile 0 to resistance profile MDR. This model is also shown in Figure \ref{fig:modelstructure}. (b) Submodel 1: no direct conversion from resistance profile 0 to resistance profile MDR $(\rho_{\mathrm{MDR}}=0)$. (c) Submodel 2: same rate of conversion for the two drugs $(\rho_{\mathrm{INH}}=\rho_{\mathrm{RIF}}=\rho_{\mathrm{single}})$. 
}
\label{fig:models}
\end{figure}

We estimate the rates of acquisition of drug resistance to rifampicin and isoniazid by fitting the model described in Section~\ref{sec:description_of_the_model_and_parameters} to the Bolivian data~\shortcite{Monteserin.13.genotypes} with the ABC method described in Section~\ref{sec:abc}. Additionally, by constraining particular resistance-acquisition parameters $\rho_k$ to produce meaningful submodels of the full model, we are able to examine two specific biological questions. The relationships between the two submodels and the full model are illustrated in Figure~\ref{fig:models}. First, we ask whether it is possible for multidrug resistance to evolve directly from doubly sensitive bacteria or whether this direct conversion does not occur (i.e., $\rho_{\rm MDR}=0$: Submodel 1). Second, we ask whether differences between rates of mutation to rifampicin and isoniazid resistance are apparent at the epidemiological scale (i.e., $\rho_{\rm INH} = \rho_{\rm RIF} = \rho_{\mathrm{single}}$: Submodel 2).

%~~~~~~~~~~~~~~~~~~~~~~~~~~~~~~~~~~~~~~~~~~~~~~~~~~~~~~~~~~~~~~~~~~~~~
%~~~~~~~~~~~~~~~~~~~~~~~~~~~~~~~~~~~~~~~~~~~~~~~~~~~~~~~~~~~~~~~~~~~~~

\begin{figure}[htp]
  \centering
  \subfloat[\footnotesize{Full model} \label{fig:marg.full}]{\includegraphics[width=5cm,height=5cm,angle=-90]{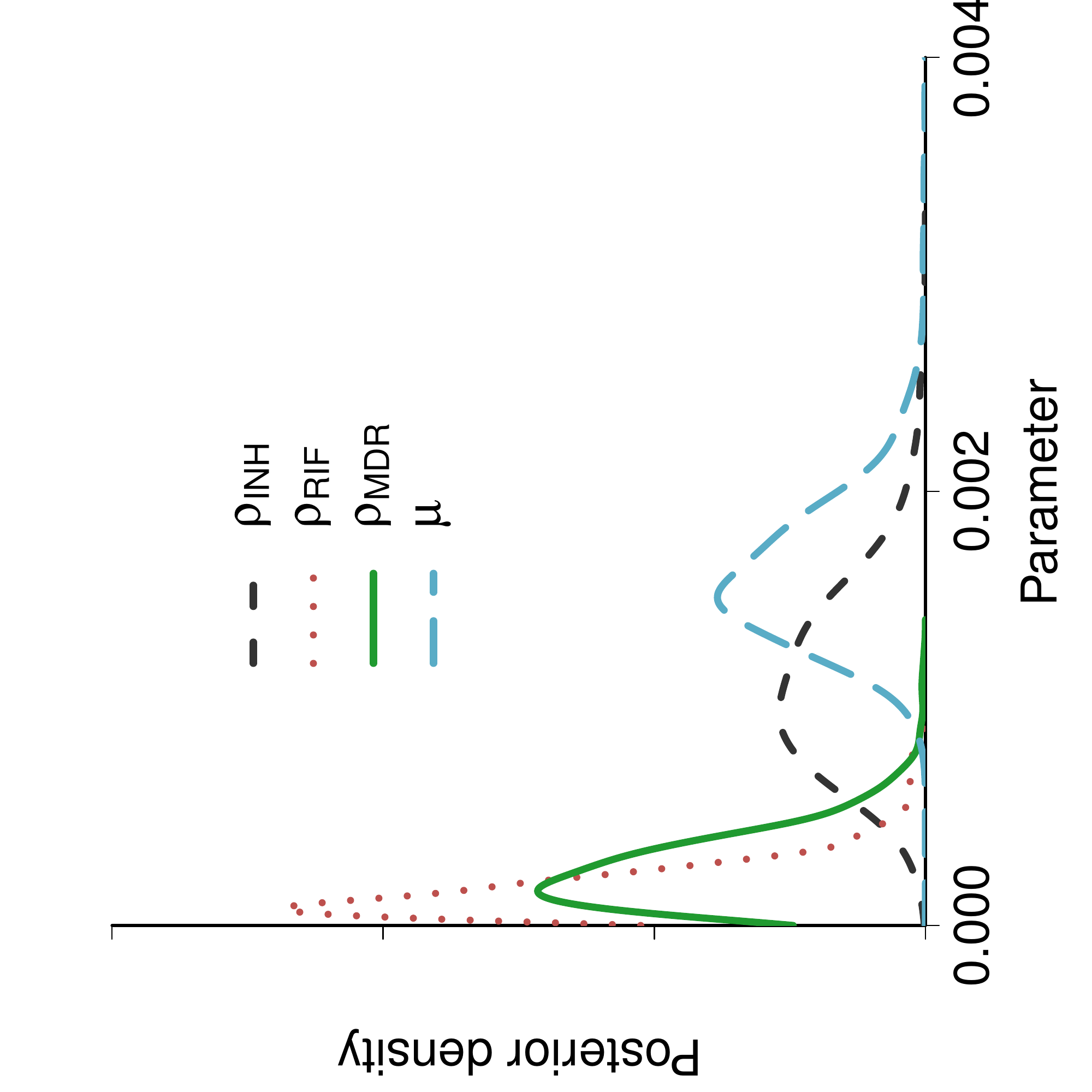}}
  \subfloat[\footnotesize{Submodel 1:} $\rho_{\mathrm{MDR}}=0$ \label{fig:marg.sub1}]{\includegraphics[width=5cm,height=5cm,angle=-90]{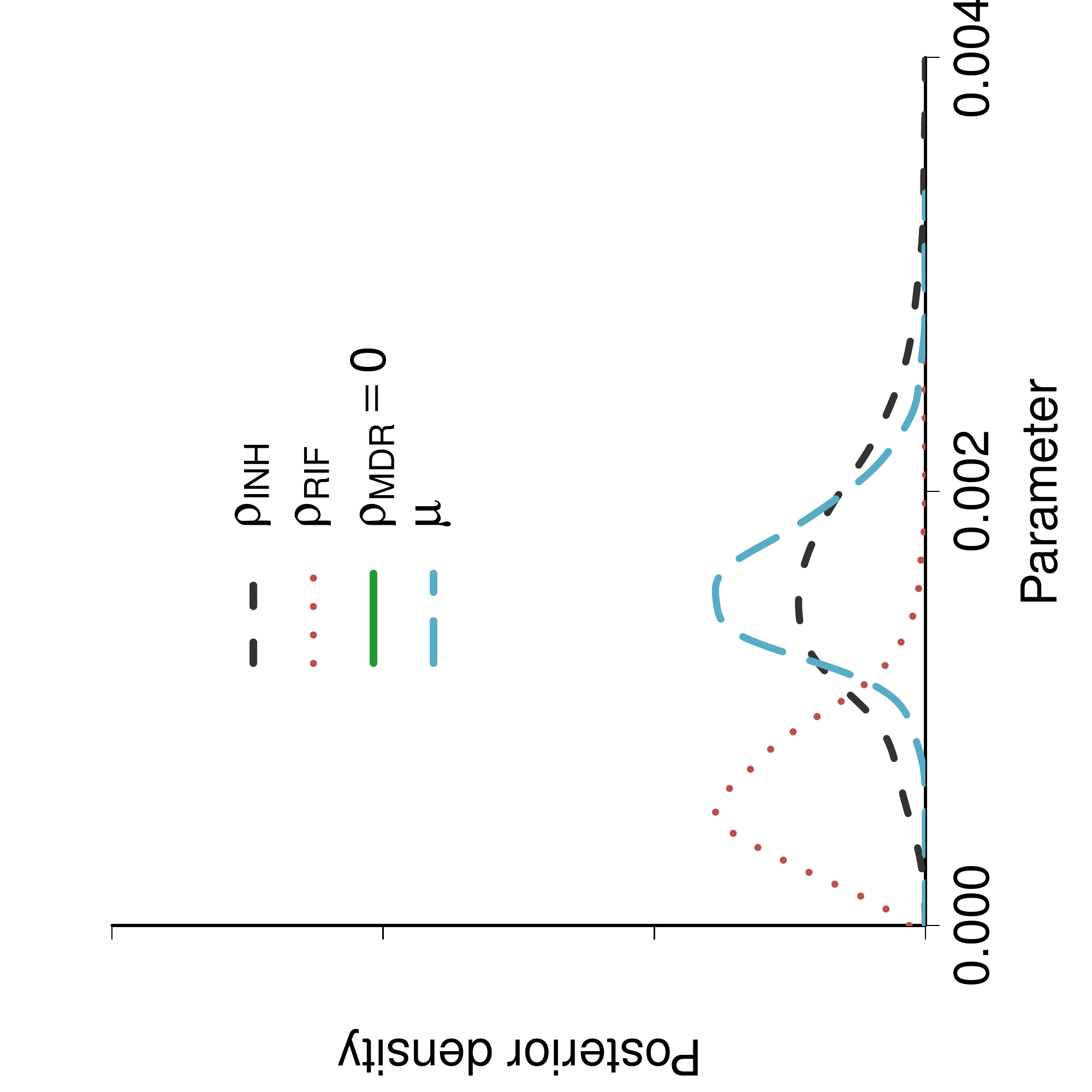}}
  \\
  \subfloat[\footnotesize{Submodel 2:} $\rho_{\mathrm{INH}}=\rho_{\mathrm{RIF}}=\rho_{\mathrm{single}}$ \label{fig:marg.sub2}]{\includegraphics[width=5cm,height=5cm,angle=-90]{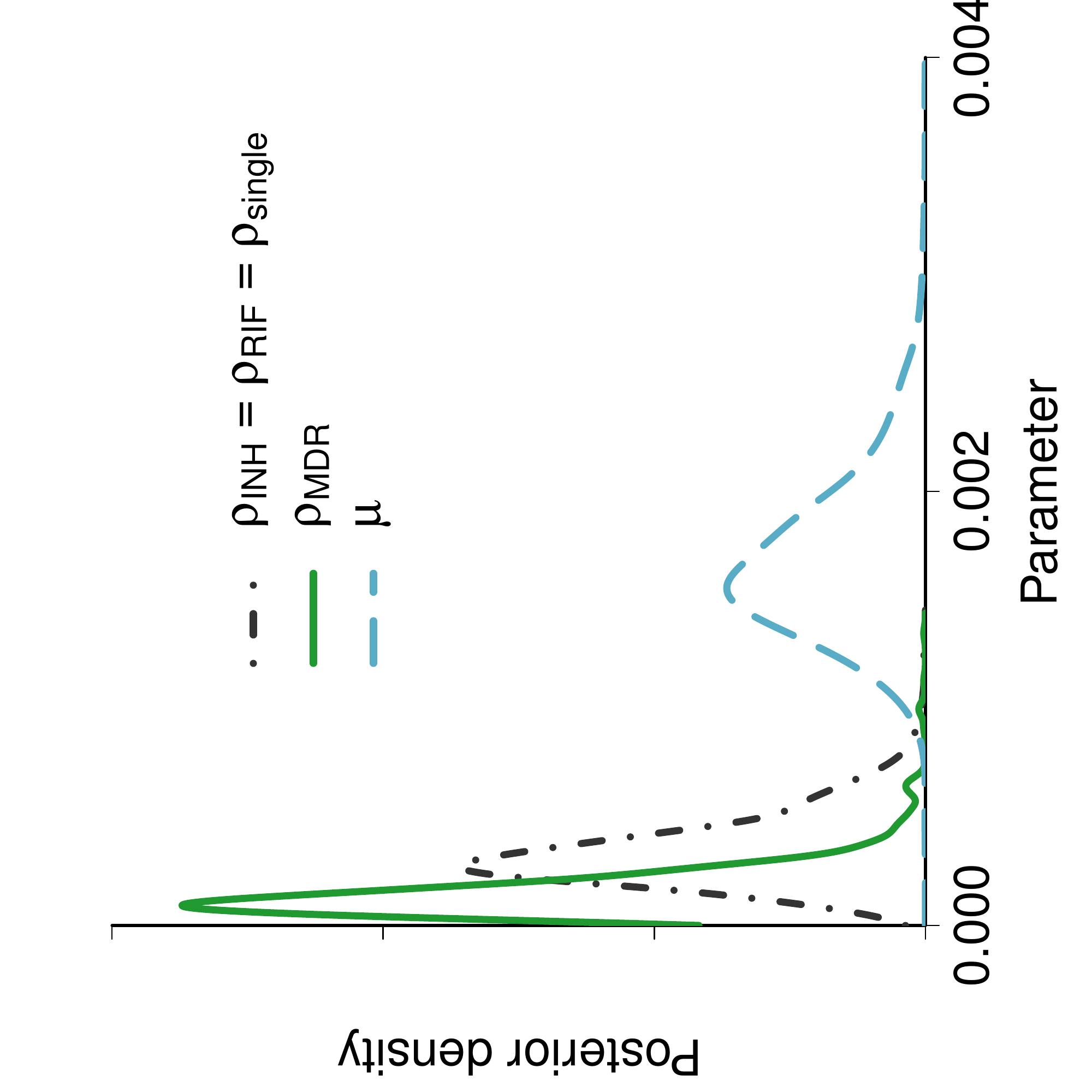}}
  \subfloat[\footnotesize{Posterior density of} $\beta_0$]{\includegraphics[width=5cm,height=5cm,angle=-90]{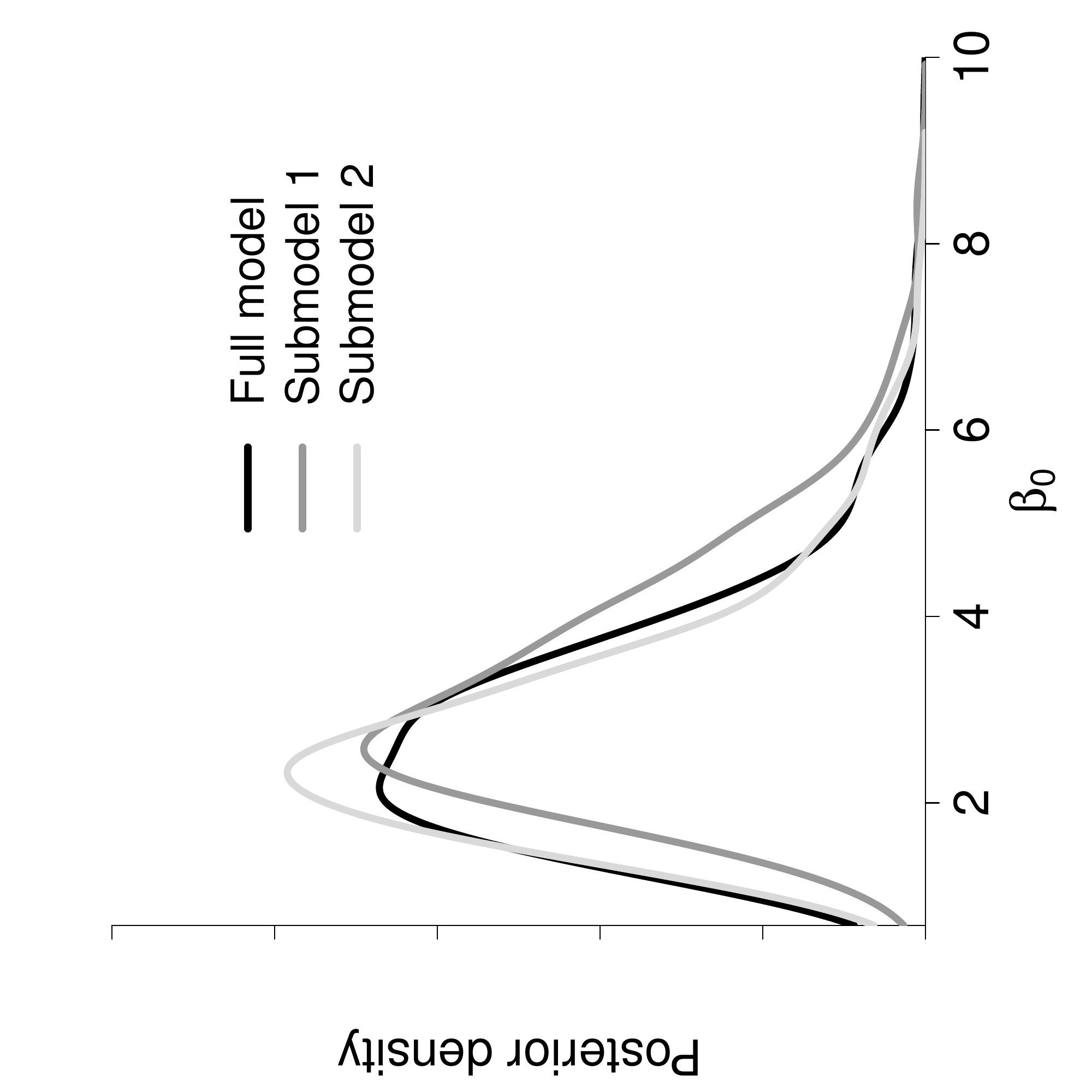}}
  \caption{\small Estimated ABC marginal posterior densities for each estimated parameter under (a) the full model, (b)  Submodel 1 ($\rho_{\mathrm{MDR}}=0$), and (c) Submodel 2 ($\rho_{\mathrm{INH}}=\rho_{\mathrm{RIF}}=\rho_{\mathrm{single}}$). Panel (d) shows the estimated ABC marginal posterior density of the transmission rate $\beta_0$ of the sensitive strain for each model structure.
  }
\label{fig:Marginals}
\end{figure}

Figure \ref{fig:Marginals} illustrates the ABC marginal posterior density estimates of each parameter under the three different sets of model assumptions. 
Under the full model there is a clear visual difference between the rates of mutation of rifampicin and isoniazid resistance, with the latter occurring at a much higher rate. In contrast, the rate of simultaneous resistance acquisition appears to be higher than that for rifampicin alone.
When eliminating the possibility of simultaneous acquisition of multiple drug resistance $\rho_{\mathrm{MDR}}=0$ (Submodel 1), $\rho_{\mathrm{INH}}$ and $\rho_{\mathrm{RIF}}$ both increase, relative to the full model, to compensate for the imposed restriction when fitting to the observed data (Figure \ref{fig:marg.sub1}). Similarly, when we fix the identity $\rho_{\mathrm{INH}}=\rho_{\mathrm{RIF}}=\rho_{\mathrm{single}}$ (Submodel 2) to impose a single rate of resistance acquisition, the posterior density of this parameter moves to intermediate values compared to the two distinct rates of acquisition estimated under the full model (Figure \ref{fig:marg.sub2}). 
The estimated posterior densities for the transmission ($\beta_0$) and mutation ($\mu$) parameters are visually similar across all models. 
ABC marginal posterior means and highest posterior density (HPD) credible intervals for all models are reported in Table \ref{tab:posterior_summaries}.

%~~~~~~~~~~~~~~~~~~~~~~~~~~~~~~~~~~~~~~~~~~~~~~~~~~~~~~~~~~~~~~~~~~~~~
\begin{table}[ht]
\centering
\begin{tabular}{lccccc}
  \vspace{.2cm}
  & $\rho_{\mathrm{INH}}$ & $\rho_{\mathrm{RIF}}$ & $\rho_{\mathrm{MDR}}$ & $\mu$ & $\beta_0$ \\ 
  \hline 
  \vspace{-.3cm}  
    \\
  \rowgroup{\bf{Full model}} \\
  Posterior mean & $1.14\times10^{-3}$ & $1.67\times10^{-4}$ & $2.62\times10^{-4}$ & $1.64\times10^{-3}$ & 2.85 \\ 
  CI lower limit & $3.40\times10^{-4}$ & $3.82\times10^{-6}$ & $3.93\times10^{-6}$ & $1.11\times10^{-3}$ & 0.97 \\ 
  \vspace{.3cm}
  CI upper limit & $1.94\times10^{-3}$ & $4.28\times10^{-4}$ & $5.81\times10^{-4}$ & $2.40\times10^{-3}$ & 5.33 \\   
    \rowgroup{\bf{Submodel l}} \\
  Posterior mean & $1.60\times10^{-3}$ & $6.37\times10^{-4}$ & -- & $1.59\times10^{-3}$ & 3.29 \\ 
  CI lower limit & $4.55\times10^{-4}$ & $1.27\times10^{-4}$ & -- & $1.03\times10^{-3}$ & 1.20 \\ 
  \vspace{.3cm}
  CI upper limit & $2.49\times10^{-3}$ & $1.24\times10^{-3}$ & -- & $2.19\times10^{-3}$ & 5.78 \\ 
  
    \rowgroup{\bf{Submodel 2}} \\
  Posterior mean & $3.46\times10^{-4}$ & -- & $1.56\times10^{-4}$ & $1.70\times10^{-3}$ & 2.81 \\ 
  CI lower limit & $7.26\times10^{-5}$ & -- & $6.62\times10^{-7}$ & $1.10\times10^{-3}$ & 0.86 \\ 
  CI upper limit & $6.90\times10^{-4}$ & -- & $3.76\times10^{-4}$ & $2.54\times10^{-3}$ & 5.20 \\ 
  
\end{tabular}
\caption{\small ABC posterior means with lower and upper limits of the 95\% HPD (highest posterior density) credibile intervals for each parameter of each fitted model.
}
\label{tab:posterior_summaries}
\end{table}
%~~~~~~~~~~~~~~~~~~~~~~~~~~~~~~~~~~~~~~~~~~~~~~~~~~~~~~~~~~~~~~~~~~~~~

%~~~~~~~~~~~~~~~~~~~~~~~~~~~~~~~~~~~~~~~~~~~~~~~~~~~~~~~~~~~~~~~~~~~~~
%~~~~~~~~~~~~~~~~~~~~~~~~~~~~~~~~~~~~~~~~~~~~~~~~~~~~~~~~~~~~~~~~~~~~~

%%%%%%%%%%%%%%%%%%%%%%%%%%%%%%%%%%%%%%%%%%
%%%%%%%%%%%%%%%%%%%%%%%%%%%%%%%%%%%%%%%%%%%
\subsection{Can resistance to both drugs be acquired simultaneously? } 
%%%%%%%%%%%%%%%%%%%%%%%%%%%%%%%%%%%%%%%%%%%
%%%%%%%%%%%%%%%%%%%%%%%%%%%%%%%%%%%%%%%%%%
\label{sec:direct_double_res}

To determine whether resistance to both drugs can evolve directly from a double sensitive strain within an infection, we compare  Submodel 1 ($\rho_{\mathrm{MDR}}=0$) against the full model.
Formal standard Bayesian model comparison typically occurs through Bayes factors. In the ABC framework this task is complicated by the need to perform ABC with summary statistics that are informative for the model indicator parameter, in addition to those informative for the model specific parameters. Such summary statistics can not only be difficult to identify, but the resulting composite vector of summary statistics can be high dimensional, which may then produce more inaccurate inference than if each model was analysed  independently. See e.g. \shortciteN{robert+cmp11}, \shortciteN{marin+prr14} and \shortciteN[this volume]{marin+pr17} for a discussion of these issues. A useful alternative is to consider posterior predictive checks or related goodness-of-fit tests (e.g. \shortciteNP{thornton+a06,csillery+bgf10,Aandahl.12.model,prangle+bps14}).

\begin{figure}[htp]
  \centering
  \subfloat[\footnotesize{Full model}]{\includegraphics[width=6cm,height=6cm,angle=-90]{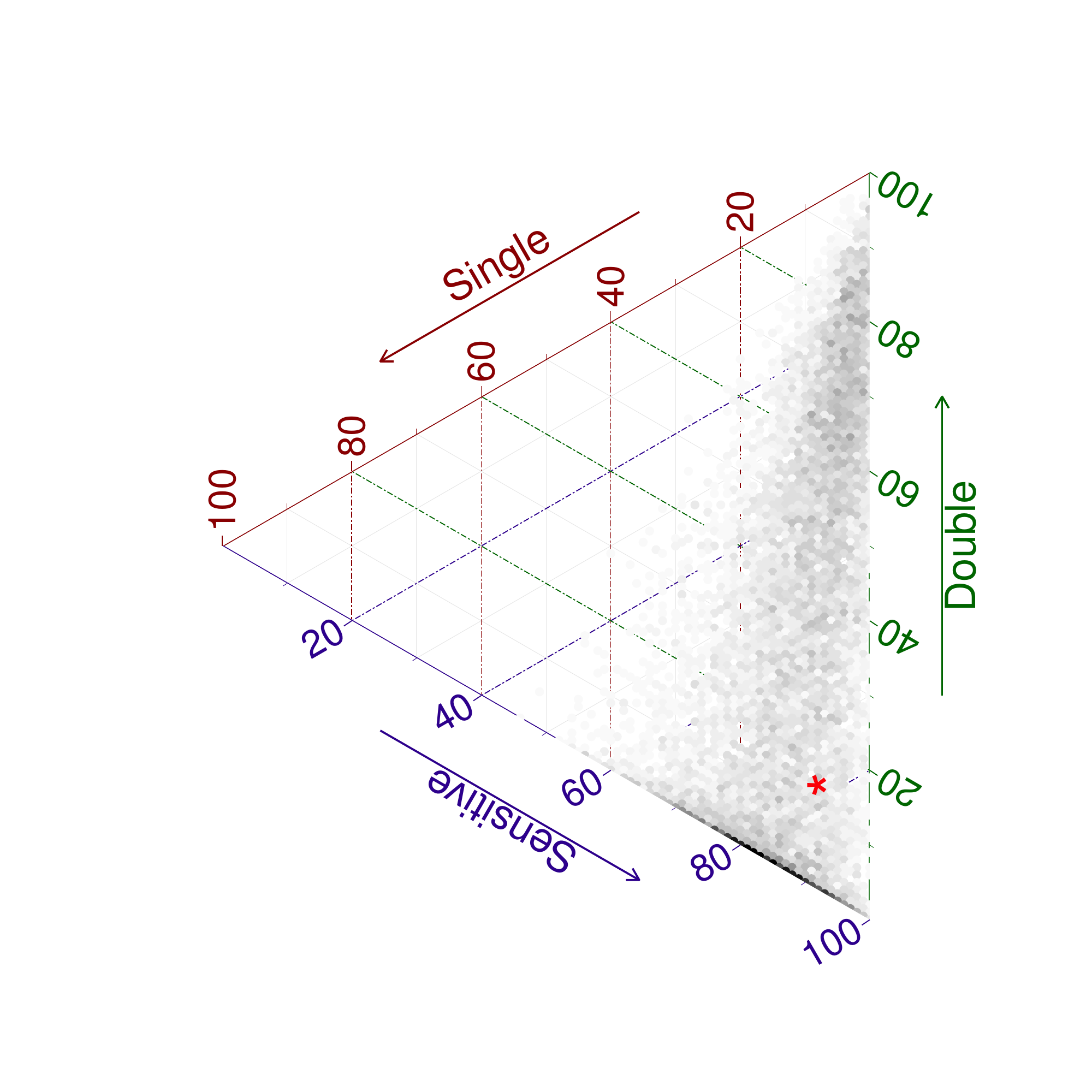}}
  \subfloat[\footnotesize{Submodel 1:} $\rho_{\mathrm{MDR}}=0$]{\includegraphics[width=6cm,height=6cm,angle=-90]{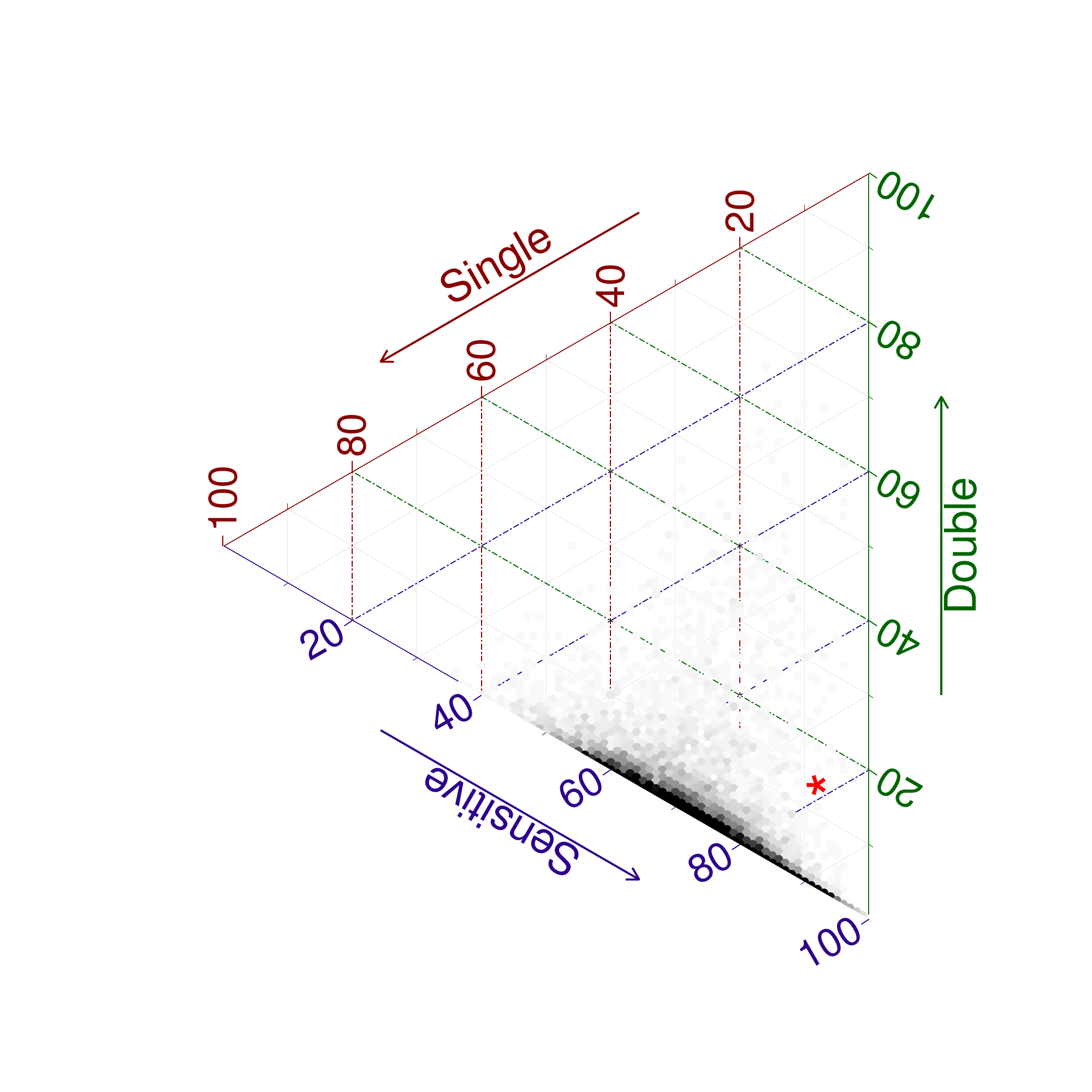}}
  \caption{\small Posterior predictive distribution of $(n_{0}, n_{\mathrm{INH}} + n_{\mathrm{RIF}}, n_{\mathrm{MDR}})$ under the full model (panel (a)) and Submodel 1 (panel (b)). Darker intensity indicates higher posterior density. The asterisk (*) indicates the observed data (78, 8, 16).
  }
  \label{fig:simplex}
\end{figure}

Figure \ref{fig:simplex} shows the posterior predictive distribution of the summary statistics $(n_{0}, n_{\mathrm{INH}} + n_{\mathrm{RIF}}, n_{\mathrm{MDR}})$ described in Section \ref{subsec:summary_statistics}, for the full model (panel (a)) and Submodel 1 (panel (b)), where a darker intensity indicates higher density.
This predictive distribution graphically illustrates each model's ability to generate the observed summary statistics (78, 8, 16), indicated by the asterisks, which represent the number of individuals in the sample sensitive to both drugs ($n_0$), resistant to a single drug ($n_{\mathrm{INH}} + n_{\mathrm{RIF}}$) and resistant to both drugs ($n_{\mathrm{MDR}}$). 
  
%~~~~~~~~~~~~~~~~~~~~~~~~~~~~~~~~~~~~~~~~~~~~~~~~~~~~~~~~~~~~~~~~~~~~~
%~~~~~~~~~~~~~~~~~~~~~~~~~~~~~~~~~~~~~~~~~~~~~~~~~~~~~~~~~~~~~~~~~~~~~
  
The predictive distributions for each model are diffuse, particularly for the full model. This variability is expected given that the sample size is small (100 isolates) and that the evolution of drug resistance from sensitivity is a relatively rare stochastic event. In the case of Submodel 1 (Figure \ref{fig:simplex} panel (b))  where we impose the condition  $\rho_{\mathrm{MDR}}=0$, the density of samples is shifted away from the bottom-right corner which represents double resistance. This pattern is due to the lack of the direct route to multidrug resistance. The observed data (asterisk) is in a region of low posterior predictive density under Submodel 1, and so we conclude that this model is not particularly supported by the data. In contrast, the observed data lie more clearly within a moderately high density region of the posterior predictive under the full model (Figure \ref{fig:simplex} panel (a)). This analysis therefore suggests that of the two competing hypotheses, it is more likely  that resistance to both drugs can be acquired simultaneously ($\rho_{\rm MDR}>0$) than otherwise. Note, however, that this direct route is not the only possible path to double resistance, which can still occur in stages through single resistance.

%%%%%%%%%%%%%%%%%%%%%%%%%%%%%%%%%%%%%%%%%%
%%%%%%%%%%%%%%%%%%%%%%%%%%%%%%%%%%%%%%%%%%%
\subsection{Is resistance to both drugs acquired at equal rates? } 
%%%%%%%%%%%%%%%%%%%%%%%%%%%%%%%%%%%%%%%%%%%
%%%%%%%%%%%%%%%%%%%%%%%%%%%%%%%%%%%%%%%%%%

\begin{figure}[htp]
  \centering
  \subfloat[\footnotesize{Full model}]{\includegraphics[width=5cm,height=5cm,angle=-90]{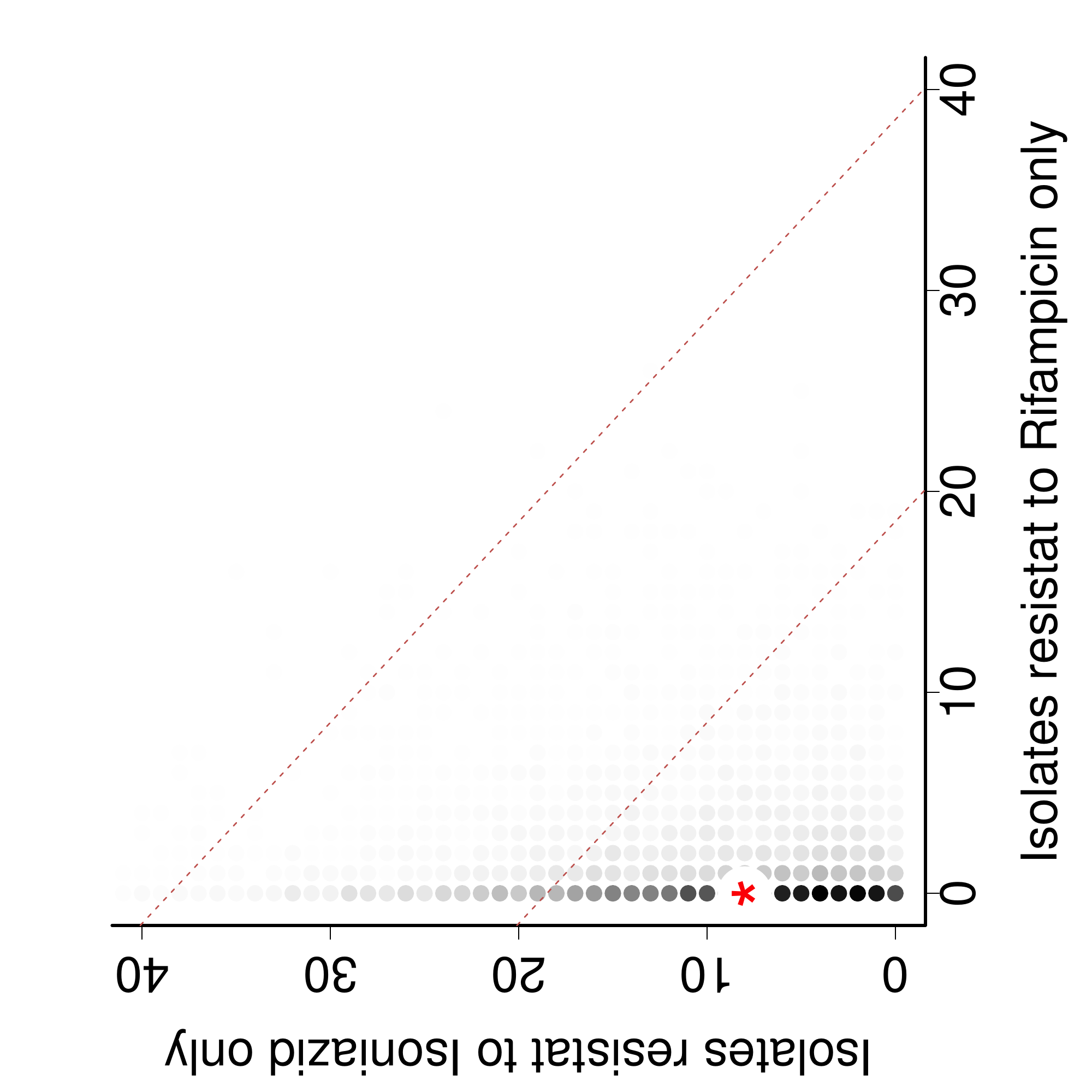}}
  \subfloat[\footnotesize{Submodel 2:} $\rho_{\mathrm{INH}}=\rho_{\mathrm{RIF}}$]{\includegraphics[width=5cm,height=5cm,angle=-90]{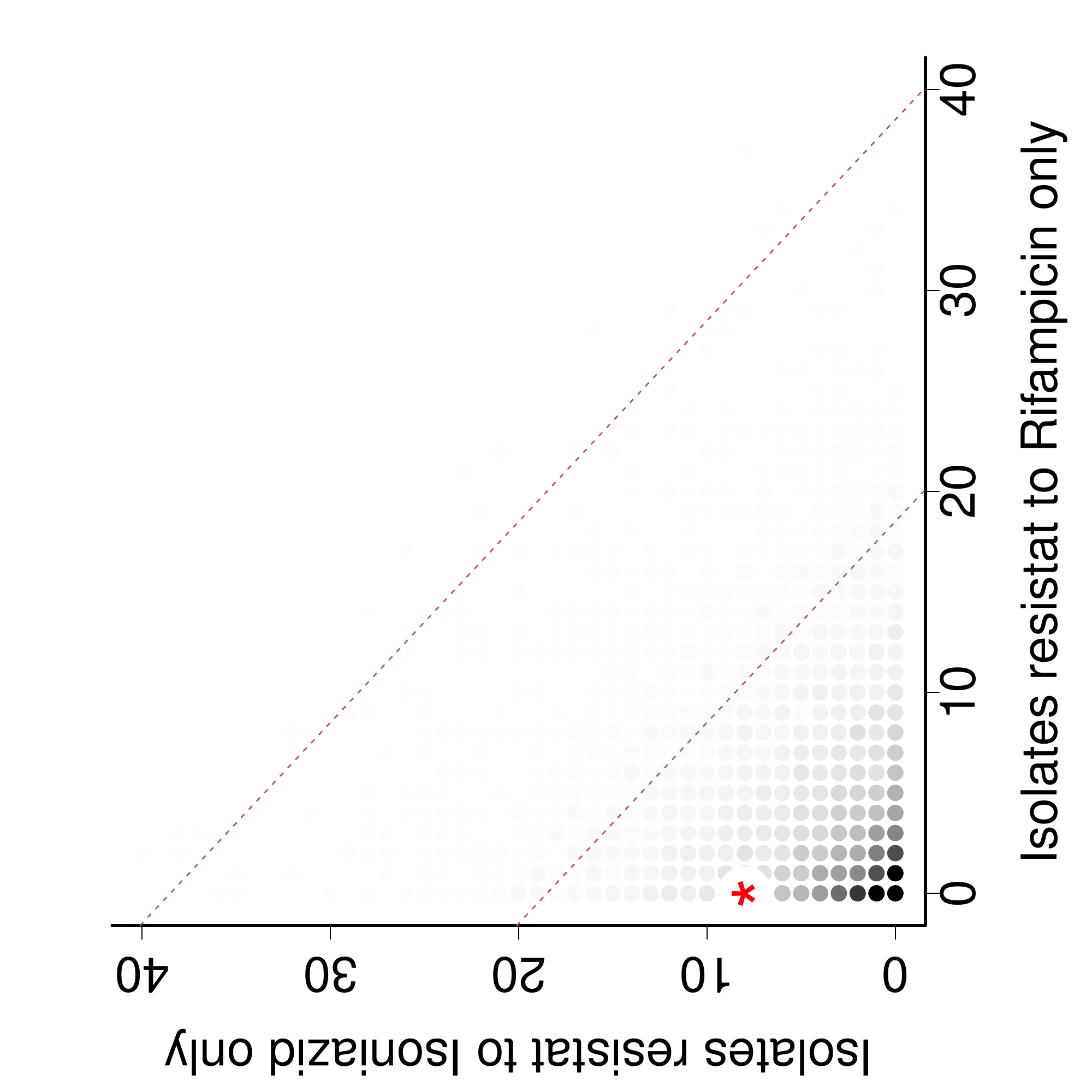}}
  \caption{\small Posterior predictive distribution of $(n_{\mathrm{INH}}, n_{\mathrm{RIF}})$ 
  under the full model (panel (a)) and Submodel 2 (panel (b)). Darker intensity indicates higher posterior predictive density. The asterisk (*) indicates the observed data (8, 0). 
}
  \label{fig:Predictive}
\end{figure}

In order to determine whether the rates of acquisition of resistance to the two drugs are equal ($\rho_{\mathrm{INH}}=\rho_{\mathrm{RIF}}$), we compare Submodel 2 against the full model. 
Figure \ref{fig:Predictive} depicts the posterior predictive distribution of $(n_{\mathrm{INH}}, n_{\mathrm{RIF}})$ under each model -- the number of cases resistant only to isoniazid ($n_{\mathrm{INH}}$) and the number of cases resistant only to rifampicin ($n_{\mathrm{RIF}}$) in the sample. The observed values of these summary statistics are $n_{\mathrm{INH}}=8$ for isoniazid and $n_{\mathrm{RIF}}=0$ for rifampicin, illustrated as the asterisk in Figure~\ref{fig:Predictive}. As Submodel 2 does not favor any drug over the other, the predictive surface is symmetric with respect to the line $n_{\mathrm{INH}} = n_{\mathrm{RIF}}$. The extra flexibility provided by the full model shifts the predictive distribution towards the observed data. While the distribution under the full model comfortably accommodates the empirical point in a high density region, the predictive distribution under Submodel 2 is much more diffuse.
This indicates that while the observed data is not unsupported under Submodel 2, it is far more likely to be observed under the full model.
As a result, we conclude that the evidence favours the drugs being acquired at different rates; specifically, isoniazid resistance evolves faster than rifampicin resistance.

%~~~~~~~~~~~~~~~~~~~~~~~~~~~~~~~~~~~~~~~~~~~~~~~~~~~~~~~~~~~~~~~~~~~~~
%~~~~~~~~~~~~~~~~~~~~~~~~~~~~~~~~~~~~~~~~~~~~~~~~~~~~~~~~~~~~~~~~~~~~~

%%%%%%%%%%%%%%%%%%%%%%%%%%%%%%%%%%%%%%%%%%
%%%%%%%%%%%%%%%%%%%%%%%%%%%%%%%%%%%%%%%%%%%
\subsection{The relative contribution of transmission and treatment failure to MDR-TB} 
%%%%%%%%%%%%%%%%%%%%%%%%%%%%%%%%%%%%%%%%%%%
%%%%%%%%%%%%%%%%%%%%%%%%%%%%%%%%%%%%%%%%%%

In addition to estimating the rates of acquisition of drug resistance and assessing whether rates differ, we may also consider where doubly resistant cases come from. That is, estimation of the relative contribution to multidrug resistant cases of transmission of existing MDR-TB strains compared to treatment failure leading to evolution of multidrug resistance. 
The posterior predicted samples generated under the full model provide a clear portrait of the  relative contribution of the different paths to achieving double resistance (see e.g. \shortciteNP{Luciani.09.epidemiological} for an additional illustration of this procedure).

Table \ref{tab:source_of_double_resistance} shows the means, medians and the 95\% HPD credible intervals for the predicted proportion of cases of double resistance from each potential source.
These proportions are obtained conditionally on there being at least one case of double resistance in the predictive sample.
Simulated samples of this nature account for 99.67\% of all predictive samples.
The predictive distributions of the proportions are highly asymmetric (not shown), making the median a more reliable point estimate than the mean.

In the overwhelming majority of posterior predictive samples, direct \emph{transmission} was the main source of acquisition of double resistance, followed by \emph{conversion} in a single step directly from a sensitive profile (from profile $0$ to MDR) and conversion in two steps via a state of resistance to a single drug (from profile $0$ to INH to MDR, or from $0$ to RIF to MDR). This analysis corroborates the finding from Section~\ref{sec:direct_double_res} that $\rho_{\mathrm{MDR}}$ is most likely positive, and furthermore that this path is likely to be of even greater importance than conversion in two steps. 

\begin{table}[ht]
\centering
\begin{tabular}{lccc}
  Source & Median & Mean & 95\% Credible Interval \\ 
  \hline
  Transmission & 0.9975 & 0.9655 & (0.7826, 0.9999) \\ 
  Conversion in one step & 0.0023 & 0.0284 & (0.0000, 0.1667) \\ 
  Conversion in two steps & 0.0000 & 0.0060 & (0.0000,  0.0073) \\ 
\end{tabular}
\caption{\small Contributions to MDR-TB from alternative sources. This table contains the posterior medians and means and lower and upper limits of the 95\% HPD credibility intervals for the proportion of double resistance cases originating from each possible source.}
\label{tab:source_of_double_resistance}
\end{table}

%%%%%%%%%%%%%%%%%%%%%%%%%%%%%%%%%%%
%%%%%%%%%%%%%%%%%%%%%%%%%%%%%%%%%%%
\section{Conclusions} 
%%%%%%%%%%%%%%%%%%%%%%%%%%%%%%%%%%%
%%%%%%%%%%%%%%%%%%%%%%%%%%%%%%%%%%%

In this chapter we have estimated epidemiological parameters describing the acquisition of multi-drug resistance in \emph{M. tuberculosis} from molecular epidemiological data~\shortcite{Monteserin.13.genotypes} using approximate Bayesian computation. The  underlying model is intended to capture essential processes that give rise to the data, namely, transmission of the disease, recovery or death, and within-host evolution giving rise to drug resistance and new genotypes at the molecular marker loci. From this analysis we may draw three major biological conclusions about the manner in which drug resistance arises.

First, there is an asymmetry in the acquisition of resistance to isoniazid and rifampicin. 
Specifically, isoniazid resistance occurs approximately an order of magnitude more frequently than resistance against rifampicin (see Table~\ref{tab:posterior_summaries}). This asymmetry in rates is consistent with \emph{in vitro} (that is, through laboratory experiments) microbiological estimates of mutation rates per cell generation which find around 1 to 2 orders of magnitude difference between the two rates \shortcite{David.70.probability,Ford.13.Mycobacterium}.

Second, the analysis supports the occurrence of \emph{direct conversion} from doubly drug sensitive to doubly resistant (MDR) infections.  
This may be initially unintuitive because under mutation alone, if mutation occurs at rate $\rho$ per gene per unit time, the rate of appearance of double mutants is $\rho^2$, which would be vanishingly small if $\rho$ is low. However, using a mathematical model, %Colijn et al~
\shortciteN{Colijn.11.Spontaneous} argued that direct conversion can occur surprisingly fast because resistant cells are sometimes present at low frequencies in a within-host population even before treatment commences. Our analysis of data at the epidemiological level is consistent with that theoretical result. 
This direct conversion to double resistance is epidemiologically important as it accelerates the accumulation of resistance, in that resistance evolution does not have to take place sequentially. Once double resistant mutants appear, transmission of these mutants further increases their prevalence in the population.

Third, the overwhelming majority of cases of multidrug resistant tuberculosis come from transmission of already multidrug resistant strains (see Table~\ref{tab:source_of_double_resistance}), a finding that is consistent with those of \shortciteN{Luciani.09.epidemiological}. 
This large contribution of transmission occurs despite the 10\% transmission cost of each resistance which results in a $\sim 20\%$  cost for MDR-TB. 
This implies that in controlling drug resistance, although there is widespread concern about treatment failure leading to rising resistance, most resistant cases 
may be
due to transmission. Therefore, although it is important to support treatment adherence, public health efforts may benefit from focusing more on preventing disease transmission. That is, control measures that reduce the incidence of new cases are likely to help reduce MDR-TB.

By developing epidemiological models with evolutionary processes we have been able to estimate parameters describing how drug resistance -- particularly multidrug resistance -- emerges in {\em M. tuberculosis}. 
Although there is existing knowledge of rates of mutation to resistant states \emph{in vitro}, there is a need to assess the extent to which those rates translate to the epidemiological level. 
Large scale molecular epidemiological models, such as those presented here, are highly complex and multidimensional, and as such, likelihood-based analyses are not straightforward mathematically or computationally. 
In such cases, approximate Bayesian computation methods present a practical and viable approach to making statistical inferences, particularly as continually advancing molecular technologies require dynamical models to be extended and refined. 
%

%%%%%%%%%%%%%%%%%
%%%%%%%%%%%%%%%%%%
\subsection*{Acknowledgements}
%%%%%%%%%%%%%%%%%%
%%%%%%%%%%%%%%%%%

GSR is funded by the CAPES Foundation via the Science Without Borders program (BEX 0974/13-7).
SAS is supported by the Australia Research Council through the Discovery Project Scheme (DP160102544). 
MMT is supported by grant DP170101917 from the Australian Research Council. 
This research includes computations using the Linux computational cluster Katana supported by the Faculty of Science, UNSW Australia.

\bibliographystyle{chicago}

\end{document}